
\input harvmac.tex

%% MACROS

% Something to deal with sub-sub-sections

\def\unlockat{\catcode`\@=11}
\def\lockat{\catcode`\@=12}

\unlockat
% Something to deal with sub-sub-sections

\def\newsec#1{\global\advance\secno by1\message{(\the\secno. #1)}
\global\subsecno=0\global\subsubsecno=0\eqnres@t\noindent
{\bf\the\secno. #1}
\writetoca{{\secsym} {#1}}\par\nobreak\medskip\nobreak}
\global\newcount\subsecno \global\subsecno=0
\def\subsec#1{\global\advance\subsecno
by1\message{(\secsym\the\subsecno. #1)}
\ifnum\lastpenalty>9000\else\bigbreak\fi\global\subsubsecno=0
\noindent{\it\secsym\the\subsecno. #1}
\writetoca{\string\quad {\secsym\the\subsecno.} {#1}}
\par\nobreak\medskip\nobreak}
\global\newcount\subsubsecno \global\subsubsecno=0
\def\subsubsec#1{\global\advance\subsubsecno by1
\message{(\secsym\the\subsecno.\the\subsubsecno. #1)}
\ifnum\lastpenalty>9000\else\bigbreak\fi
\noindent\quad{\secsym\the\subsecno.\the\subsubsecno.}{#1}
\writetoca{\string\qquad{\secsym\the\subsecno.\the\subsubsecno.}{#1}}
\par\nobreak\medskip\nobreak}

\def\subsubseclab#1{\DefWarn#1\xdef
#1{\noexpand\hyperref{}{subsubsection}%
{\secsym\the\subsecno.\the\subsubsecno}%
{\secsym\the\subsecno.\the\subsubsecno}}%
\writedef{#1\leftbracket#1}\wrlabeL{#1=#1}}% Macros for boxes
\lockat

\def\IL{\relax{\rm I\kern-.18em L}}
\def\IH{\relax{\rm I\kern-.18em H}}
\def\IR{\relax{\rm I\kern-.18em R}}
\def\IC{\relax\hbox{$\inbar\kern-.3em{\rm C}$}}
\def\IZ{\relax\ifmmode\mathchoice
{\hbox{\cmss Z\kern-.4em Z}}{\hbox{\cmss Z\kern-.4em Z}}
{\lower.9pt\hbox{\cmsss Z\kern-.4em Z}}
{\lower1.2pt\hbox{\cmsss Z\kern-.4em Z}}\else{\cmss Z\kern-.4em
Z}\fi}
\def\CM {{\cal M}}
\def\CN {{\cal N}}

\def\CD {{\cal D}}
\def\CF {{\cal F}}

\def\CO {{\cal O}}

\def\CB {{\cal B}}

\def\CA{{\cal A}}

%% MORE MACROS
\def\CM {{\cal M}}
\def\CN {{\cal N}}

\def\CO {{\cal O}}

\def\CQ {{\cal Q }}

\font\manual=manfnt \def\dbend{\lower3.5pt\hbox{\manual\char127}}

\def\IZ{\relax\ifmmode\mathchoice
{\hbox{\cmss Z\kern-.4em Z}}{\hbox{\cmss Z\kern-.4em Z}}
{\lower.9pt\hbox{\cmsss Z\kern-.4em Z}}
{\lower1.2pt\hbox{\cmsss Z\kern-.4em Z}}\else{\cmss Z\kern-.4em
Z}\fi}

\def\p{\partial}

\def\CM {{\cal M}}
\def\CN {{\cal N}}

\def\CO {{\cal O}}

\def\CQ {{\cal Q }}

% more macros, alphabetically

\def\IZ{\relax\ifmmode\mathchoice
{\hbox{\cmss Z\kern-.4em Z}}{\hbox{\cmss Z\kern-.4em Z}}
{\lower.9pt\hbox{\cmsss Z\kern-.4em Z}}
{\lower1.2pt\hbox{\cmsss Z\kern-.4em Z}}\else{\cmss Z\kern-.4em
Z}\fi}
\def\IB{\relax{\rm I\kern-.18em B}}
\def\IC{{\relax\hbox{$\inbar\kern-.3em{\rm C}$}}}
\def\ID{\relax{\rm I\kern-.18em D}}
\def\IE{\relax{\rm I\kern-.18em E}}
\def\IF{\relax{\rm I\kern-.18em F}}
\def\IG{\relax\hbox{$\inbar\kern-.3em{\rm G}$}}
\def\IGa{\relax\hbox{${\rm I}\kern-.18em\Gamma$}}
\def\IH{\relax{\rm I\kern-.18em H}}
\def\II{\relax{\rm I\kern-.18em I}}
\def\IK{\relax{\rm I\kern-.18em K}}
\def\IP{\relax{\rm I\kern-.18em P}}
\def\IQ{\relax\hbox{$\inbar\kern-.3em{\rm Q}$}}

\def\liet{{\underline{\bf t}}}

\def\inbar{\,\vrule height1.5ex width.4pt depth0pt}

\def\mod{{\rm mod}}
\def\p{\partial}

\font\cmss=cmss10 \font\cmsss=cmss10 at 7pt
\def\IR{\relax{\rm I\kern-.18em R}}

\def\Tr{\rm Tr}

% Macros for boxes

\def\boxit#1{\vbox{\hrule\hbox{\vrule\kern8pt
\vbox{\hbox{\kern8pt}\hbox{\vbox{#1}}\hbox{\kern8pt}}
\kern8pt\vrule}\hrule}}
\def\mathboxit#1{\vbox{\hrule\hbox{\vrule\kern8pt\vbox{\kern8pt
\hbox{$\displaystyle #1$}\kern8pt}\kern8pt\vrule}\hrule}}

%% ANOTHER SET OF MACROS

\def\liet{{\underline{\bf t}}}

\def\inbar{\,\vrule height1.5ex width.4pt depth0pt}

\def\p{\partial}

\font\cmss=cmss10 \font\cmsss=cmss10 at 7pt
\def\IR{\relax{\rm I\kern-.18em R}}

\def\rank{{\rm rank}}

\def\Tr{\rm Tr}

%% new macros

%% END MACROS
%%

\lref\bateman{A. Erdelyi et. al. , {\it Higher
Transcendental Functions, vol. I,
Bateman manuscript project} (1953)
McGraw-Hill}

\lref\borchaut{R. Borcherds,
``Automorphic forms with singularities on
Grassmannians,'' alg-geom/9609022}

\lref\cohen{H. Cohen, ``Sums involving the
values at negative integers of L-functions of
quadratic characters,''
Math. Ann. {\bf 217}(1975) 271}

\lref\donint{S.K. Donaldson, ``Connections,
Cohomology and the intersection forms of
4-manifolds,'' J. Diff. Geom. {\bf 24 }
(1986)275.}

\lref\eg{G. Ellingsrud and L. G\"ottsche,
``Wall-crossing formulas, Bott residue formula and
the Donaldson invariants of rational surfaces,''
alg-geom/9506019.}

\lref\ez{M. Eichler and D. Zagier, {\it The theory of
Jacobi forms}, Birkh\"auser, 1985}

\lref\finstern{R. Fintushel and R.J. Stern,
``The blowup formula for Donaldson invariants,''
alg-geom/9405002; Ann. Math. {\bf 143} (1996) 529.}

\lref\gottsche{L. G\"ottsche, ``Modular forms and Donaldson
invariants for 4-manifolds with $b_+=1$,'' alg-geom/9506018; J. Am. Math. Soc.
{\bf 9}
(1996) 827.}

\lref\gottzag{L. G\"ottsche and D. Zagier,
``Jacobi forms and the structure of Donaldson
invariants for 4-manifolds with $b_+=1$,''
alg-geom/9612020.}

\lref\mw{G. Moore and E. Witten, ``Integration over
the $u$-plane in Donaldson theory," hep-th/9709193.}

\lref\swi{N. Seiberg and E. Witten,
``Electric-magnetic duality, monopole condensation, and confinement in
${\cal N}=2$ supersymmetric Yang-Mills
Theory,''
hep-th/9407087; Nucl. Phys. {\bf B426} (1994) 19}

\lref\swii{N. Seiberg and E. Witten,
``Monopoles, Duality and Chiral Symmetry Breaking in N=2 Supersymmetric QCD,''
hep-th/9408099}

\lref\klemmrev{A. Klemm,
``On the geometry behind ${\cal N}=2$ supersymmetric effective actions in four
dimensions,''
hep-th/9705131}

\lref\klty{A. Klemm, W. Lerche, S. Theisen and S. Yankielowicz, ``Simple
singularities and ${\cal N}=2$ supersymmetric Yang-Mills theory,"
hep-th/9411048;
Phys. Lett. {\bf B344} (1995) 169. }

\lref\af{ P.C. Argyres and A.E. Faraggi, ``Vacuum structure and spectrum
of ${\cal N}=2$ supersymmetric $SU(N)$ gauge theory," hep-th/9411057; Phys.
Rev.
Lett.
{\bf 74} (1995) 3931.}

\lref\kltytwo{A. Klemm, W. Lerche, S. Theisen and S. Yankielowicz, ``On the
monodromies of ${\cal N}=2$ supersymmetric Yang-Mills theory," hep-th/9412158.}

\lref\klt{A. Klemm, W. Lerche and S. Theisen, ``Nonperturbative effective
actions of ${\cal N}=2$ supersymmetric gauge theories," hep-th/9505150; Int. J.
Mod.
Phys. {\bf A10} (1996) 1929.}

\lref\ad{P.C. Argyres and M.R. Douglas, ``New phenomena in $SU(3)$
supersymmetric
gauge theory," hep-th/9505062; Nucl. Phys. {\bf B448} (1995) 93.}
\lref\ds{M.R. Douglas and S.H. Shenker, ``Dynamics of $SU(N)$ supersymmetric
gauge theory,"
hep-th/9503163; Nucl. Phys. {\bf B447} (1995) 271.}

\lref\giveonkut{A. Giveon and D. Kutasov,
``Brane Dynamics and Gauge Theory,'' hep-th/9802067}

\lref\vw{C. Vafa and E. Witten,
``A strong coupling test of $S$-duality,''
hep-th/9408074; Nucl. Phys. {\bf B431} (1994) 3.}

\lref\monopole{E. Witten, ``Monopoles and
four-manifolds,''  hep-th/9411102; Math. Res. Letters {\bf 1} (1994) 769.}

\lref\witteni{E. Witten, ``On $S$-duality in abelian
gauge theory,'' hep-th/9505186; Selecta Mathematica {\bf 1} 383}

\lref\wittk{E. Witten, ``Supersymmetric Yang-Mills theory
on a four-manifold,''  hep-th/9403193;
J. Math. Phys. {\bf 35} (1994) 5101.}

\lref\zagi{D. Zagier, ``Nombres de classes et formes
modulaires de poids 3/2,'' C.R. Acad. Sc. Paris,
{\bf 281A} (1975)883.}
\lref\zagii{F. Hirzebruch and D. Zagier,
``Intersection numbers of curves on Hilbert modular
surfaces and modular forms of Nebentypus,''
Inv. Math. {\bf 36}(1976)57.}

\lref\lns{A. Losev, N. Nekrasov, and S. Shatashvili, ``Issues in
topological gauge theory," hep-th/9711108; ``Testing Seiberg-Witten solution,"
hep-th/9801061.}

\lref\ms{T. Masuda and H. Suzuki, ``On explicit evaluations around the
conformal point in ${\cal N}=2$ supersymmetric Yang-Mills theories,"
hep-th/9612240; Nucl. Phys. {\bf B495} (1997) 149.}

\lref\mm{M. Mari\~no and G. Moore, ``Integrating over the Coulomb branch in
${\cal N}=2$ gauge theory," hep-th/9712062.}

\lref\matone{M. Matone, ``Instantons and recursion relations in ${\cal N}=2$
supersymmetric
gauge theory," hep-th/9506102; Phys. Lett. {\bf B357} (1995) 342.}

\lref\ey{T. Eguchi and S.K. Yang, ``Prepotentials of ${\cal N}=2$
supersymmetric gauge
theories
and soliton equations," hep-th/9510183; Mod. Phys. Lett. {\bf A11} (1996) 131.}

\lref\sty{J. Sonnenschein, S. Theisen, and S. Yankielowicz, ``On the relation
between the holomorphic prepotential and the quantum moduli in supersymmetric
gauge theories,"
hep-th/9510129; Phys. Lett. {\bf B367} (1996) 145.}

\lref\ky{T. Kubota and N. Yokoi, ``Renormalization group flow near the
superconformal
points in ${\cal N}=2$ supersymmetric gauge theories," hep-th/9712054.}

\lref\humphreys{J.E. Humphreys, {\it Introduction to Lie algebras and
representation theory},
Springer-Verlag, 1972.}

\lref\mor{A. Gorsky, I.M. Krichever, A. Marshakov, A. Mironov and A.Morozov,
``Integrability
and Seiberg-Witten exact solution," hep-th/9505035; Phys. Lett. {\bf B355}
(1995) 466.}

\lref\marwar{E. Martinec and N.P. Warner, ``Integrable systems and
supersymmetric gauge theory," hep-th/9509161; Nucl. Phys. {\bf B459} (1996) 97.
}

\lref\dw{R. Donagi and E. Witten, ``Supersymmetric Yang-Mills theory and
integrable systems",
hep-th/9510101; Nucl. Phys. {\bf B460} (1996) 299.}

\lref\don{R. Donagi, alg-geom/9705010.}

\lref\wittmfive{E. Witten, ``Solutions of four-dimensional field theories
via M-theory,"  hep-th/9703066; Nucl. Phys. {\bf B500} (1997) 3.}

\lref\fandm{R. Friedman and J.W. Morgan,
``Algebraic surfaces and Seiberg-Witten invariants,''
alg-geom/9502026; J. Alg. Geom. {\bf 6} (1997) 445. ``Obstruction bundles,
semiregularity, and Seiberg-Witten
invariants'', alg-geom/9509007.}

\lref\morganbk{J.W. Morgan, {\it The Seiberg-Witten equations and applications
to the topology of smooth four-manifolds}, Princeton University Press, 1996.}

\lref\DoKro{S.K.~ Donaldson and P.B.~ Kronheimer,
{\it The Geometry of Four-Manifolds},
Clarendon Press, Oxford, 1990.}

\lref\FrMor{R. Friedman and J.W. Morgan,
{\it Smooth Four-Manifolds and Complex Surfaces},
Springer Verlag, 1991.}

\lref\tqft{E. Witten,
``Topological Quantum Field Theory,''
Commun. Math. Phys. {\bf 117} (1988)
353.}
\lref\dansun{U.H. Danielsson and B. Sundborg, ``The moduli space and
monodromies of
${\cal N}=2$ supersymmetric gauge theories," hep-th/9504102; Phys. Lett. {\bf
B358}
(1995) 273.}

\lref\naka{T. Nakatsu and K. Takasaki, ``Whitham-Toda hierarchy and ${\cal
N}=2$
supersymmetric
Yang-Mills theory," hep-th/9509162; Mod. Phys. Lett. {\bf A11} (1996) 157.}

\lref\gorsky{A. Gorsky, A. Marshakov, A. Mironov and A.Morozov, ``RG equations
from
Whitham hierarchy," hep-th/9802007.}
\lref\apsw{P.C. Argyres, M.R. Plesser, N. Seiberg, and E. Witten, ``New ${\cal
N}=2$ superconformal field theories in four dimensions," hep-th/9511154; Nucl.
Phys. {\bf B461} (1996)
71.}

\lref\eguchisc{T. Eguchi, K. Hori, K. Ito, and S.K. Yang, ``Study of ${\cal
N}=2$ superconformal
field theories in four dimensions," hep-th/9603002; Nucl. Phys. {\bf B471}
(1996) 430.}

\lref\dpstrong{E. D'Hoker and D.H. Phong, ``Strong coupling expansions of
$SU(N)$ Seiberg-Witten theory," hep-th/9701055; Phys. Lett. {\bf B397} (1997)
94. }

\lref\BSV{M. Bershadsky, V. Sadov, and
C. Vafa, ``D-Branes and Topological Field
Theories,'' Nucl. Phys. {\bf B463} (1996) 420; hep-th/9511222.}

\lref\susskind{T.Banks,  W.Fischler ,  I.R.Klebanov,  L.Susskind,
``Schwarzschild black holes from Matrix theory,''  hep-th/9709108}

\lref\nsc{M. Mari\~no and G. Moore, ``Donaldson invariants for nonsimply
connected manifolds," hep-th/9804104.}

\lref\doncobord{S. Donaldson,
``Irrationality and the $h$-cobordism conjecture,''
J. Diff. Geom. {\bf 26} (1987) 141.}

\lref\freedman{D. Anselmi, J. Erlich, D.Z. Freedman,
and A.A. Johansen, ``Nonperturbative formulas for
central functions of supersymmetric gauge theories,''
hep-th/9708042;
``Positivity constraints on anomalies in supersymmetric gauge theories,''
hep-th/9711035}

\lref\mmprep{M. Mari\~no and G. Moore, work in
progress.}

\lref \pidtyurin{V. Pidstrigach and A. Tyurin, ``Localization of the Donaldson
invariants along Seiberg-Witten classes,'' dg-ga/9507004.}

\lref\feehan{P.M.N. Feehan and  T.G. Leness, ``$PU(2)$ monopoles and relations
between four-manifold invariants," dg-ga/9709022; ``$PU(2)$ monopoles I:
Regularity,
Uhlenbeck compactness, and transversality," dg-ga/9710032; ``$PU(2)$ monopoles
II:
Highest-level singularities and realtions between four-manifold invariants,"
dg-ga/9712005. }

\lref\holography{ E. Witten, ``Anti de Sitter space and holography",
hep-th/9802150.}

\lref\icm{G. Moore, ``2D Yang-Mills Theory and Topological Field Theory,''
hep-th/9409044; and in
{Proceedings of the International Congress of Mathematicians 1994,}
Birkh\"auser 1995 }

\lref\gv{R. Gopakumar and C. Vafa, ``Topological gravity from large $N$ gauge
theory,"
hep-th/9802016.}

%%%%%%%%%%
\Title{\vbox{\baselineskip12pt
\hbox{YCTP-P3-98 }
\hbox{hep-th/9802185}
}}
{\vbox{\centerline{The Donaldson-Witten function }
\centerline{ }
\centerline{ for   gauge groups  }
\centerline{ }
\centerline{of rank larger than one }}
}
\centerline{Marcos Mari\~no and Gregory Moore}

\bigskip
{\vbox{\centerline{\sl Department of Physics, Yale University}
\vskip2pt
\centerline{\sl New Haven, CT 06520}}
\centerline{ \it marino@genesis5.physics.yale.edu }
\centerline{ \it moore@castalia.physics.yale.edu }

\bigskip
\bigskip
\noindent
We study correlation functions in topologically
twisted $\CN=2, d=4$ supersymmetric
Yang-Mills theory for gauge groups of rank
larger than one on compact four-manifolds
$X$.  We find that the topological invariance
of the generator of correlation functions of
BRST invariant observables is  not spoiled
by noncompactness of field space.
We show how to express the correlators
on simply connected manifolds of
$b_{2,+}(X)>0$ in terms of Seiberg-Witten invariants
and the classical cohomology ring of $X$.
For manifolds $X$ of simple type and
gauge group $SU(N)$ we give
explicit expressions of the correlators
as a sum over $\CN=1$ vacua.
We describe two applications of our expressions,
one to superconformal field theory and one to
large $N$ expansions of $SU(N)$
$\CN=2, d=4$ supersymmetric
Yang-Mills theory.

\Date{February 21, 1998}

%\draft

\newsec{Introduction and conclusion }

The Donaldson invariants of 4-manifolds
have played an important role in the
development of both mathematics and
physics  during  the past fifteen  years.
Donaldson's invariants are defined using
nonabelian gauge theory for gauge
group $G= SU(2)$ or  $G= SO(3)$
on a compact oriented Riemannian
4-manifold $X$ \DoKro\FrMor.
They were interpreted
by Witten in \tqft\ as correlation functions in
an $\CN=2$ supersymmetric Yang-Mills theory
(SYM)
and as such are best presented as
a function on $H_*(X)$ defined by
a path integral:
\eqn\dwfunction{
Z_{DW}(v\cdot S + p P) =
\biggl\langle \exp[ v\cdot I(S) + p \CO(P)] \biggr\rangle
}
where $P\in H_0(X;\IZ)$, $S  \in H_2(X;\IZ)$,
$I(S)$ and $\CO(P)$ are certain operators
in the gauge theory, and the right hand
side of \dwfunction\  is an expectation value.
We refer to \dwfunction\ as the
Donaldson-Witten function. Witten's
interpretation has lead to significant
progress in the subject \monopole.

\subsec{Questions, and answers.}

Since \dwfunction\ is a correlation function
in an $SU(2)$ or $SO(3)$
gauge theory it is quite natural
to ask about the generalization to
compact simple gauge groups $G$
of rank   larger than one.  The formal
definitions, both mathematical and physical,
proceed with little essential modification to
the higher rank case so we may ask the
following  three basic questions:

\item{1.} Is $Z_{DW}$ an invariant of the
diffeomorphism type of $X$?

\item{2.} Does $Z_{DW}$ define new 4-manifold
invariants that go beyond the classical cohomology
ring and the Seiberg-Witten invariants?

\item{3.} Can $Z_{DW}$ be evaluated explicitly?

In this paper we answer these questions:

\item{1.} Yes, $Z_{DW}$ is a topological invariant for
$\rank(G) \equiv r > 1$.

\item{2.}  No, $Z_{DW}$ does not
contain any new topological information,
at least for 4-manifolds of $b_{2,+}>0$.

\item{3.} Yes, $Z_{DW}$ can be explicitly evaluated
in terms of the classical cohomology ring and
Seiberg-Witten invariants.

These conclusions require further comment.

The first question is not silly. From the
mathematical point of view the instanton
moduli spaces are quite singular and it is
not obvious that there is a well-defined
intersection theory on them. From the physical point
of view, although the
path integral is formally topologically invariant
the expression for $Z_{DW}$ given below is,
to say the
least, intricate and delicate, and involves
integrals over noncompact spaces with
singular integrands. Given the phenomenon
of wall-crossing
\doncobord\eg\gottsche\gottzag\
and the surprising discovery
of \mw\ of continuous metric dependence in
a superconformal $SU(2)$ theory the topological
invariance of $Z_{DW}$ is not obvious. Among other things one
should worry about continuous metric
dependence of $Z_{DW}$ arising from
integration over the subvarieties in the moduli
space of supersymmetric vacua with superconformal
symmetry. The main technical work
in this paper consists of carefully
defining the integrals
and checking their metric dependence.
Our conclusion, as stated, is that there is
no continuous variation.
Somewhat surprisingly, in stark contrast to
the rank one case, we find that there is
no wall-crossing from the measure in
the  semiclassical regime.

The answer to the second question is, of course,
a disappointment. One of the main motivations
for this work was the suggestion of E. Witten,
made during the investigations of  \mw, that
wall-crossing phenomena at superconformal
points would lead to the discovery of new 4-manifold
invariants. We would like to stress that we are
{\it not} suggesting that $\CN=2$ superconformal
theories provide no new topological information
(in fact we believe the opposite).
However, if there are new invariants, they are
inaccessible via the wall-crossing technique
used in \mw\ and described below.

Regarding the third question,
the general formula is rather complicated
and is only described in full detail for
$G= SU(3)$ in equations $(9.1)-(9.6)$
 below. An important
representative case is that of simply connected
manifolds of $b_{2,+}>1$ and of simple type.
The resulting expression for $G=SU(N)$ is
given in equation $(9.13)$ below. It is a natural
generalization of the expression found by
Witten in \monopole\ for the rank one case.

\subsec{Method of derivation}

Deriving the higher rank Donaldson invariants using
the standard mathematical
methods of \DoKro\FrMor\ does not work very well.
Formal aspects of the problem, like the
$\mu$-map generalize straightforwardly but,
because of the singularities of instanton moduli space,
the intersection theory is difficult to define.

The physical approach to the problem
turns out to be much more powerful.
By the physical approach we mean
the program proposed by Witten in \tqft,
and brought to fruition in \monopole.
Some further technical developments
described in \mw\  make the derivation of
the main result of \monopole\ and its
higher rank generalizations conceptually
straightforward, (although technically
challenging in the higher rank case).

The main insight we use from \mw\ is that
one can derive the relation between Donaldson
and Seiberg-Witten invariants from the
phenomenon of wall-crossing.
\foot{Some mathematical papers dealing with the
relation of Seiberg-Witten and Donaldson invariants
are \pidtyurin\feehan.} This wall-crossing
method proceeds as follows. One
begins by considering  the contribution
to $Z_{DW}$
 of the
Coulomb branch of the moduli
space of supersymmetric  vacua
on $\IR^4$. This contribution,
denoted
by $Z_{\rm Coulomb} $,  is nonzero only for manifolds
of $b_{2,+}=1$. Nevertheless $Z_{\rm Coulomb} $
turns out to contain the essential
information for deriving the contributions of the
SW branch to $Z_{DW}$.
 In particular, cancellation of metric-dependence
of $Z_{\rm Coulomb} $
at   strong coupling singularities in moduli space
allows a
complete derivation of the universal functions
appearing in the Lagrangian of the magnetic
dual theory with the light monopole or dyon
hypermultiplet fields included in the theory.
(See section  seven of \mw.)

The wall-crossing method generalizes to
simple gauge groups of rank $r>1$.
The Coulomb
branch $\CM_{\rm Coulomb} $
 is now a quasi-affine variety of complex
dimension $r$. In the weak coupling
asymptotic region  $\CM_{\rm Coulomb} $
may be described as $(\liet \otimes \IC)/W$ where
$\liet$ is a Cartan subalgebra for $G$ and $W$ is the
Weyl group. More globally, the Coulomb branch
has the form:
\eqn\coulbrnch{
\CM_{\rm Coulomb}  = \bigl( \IC^r - \CD)
}
In Seiberg-Witten theory the space of vacuum
expectation values $\langle \Tr \phi^j \rangle $,  for $j$
ranging over the the exponents of $G$,
is identified with   $\IC^r$. The low energy theory is
characterized by a family of Abelian varieties over
$\IC^r$, and $\CD$ is the singular locus for this family.
One can introduce local special coordinates $a^I$ on
\coulbrnch, but these are never global coordinates
and together with their duals $a_{D,I}$  transform
in nontrivial ways under   the quantum monodromy
group $\Gamma$, determined in principle from the
explicit SW curve and differential.
For the example of  $SU(N)$,  $\CD$ is defined
by the vanishing of the ``quantum discriminant''
of equation $(2.10)$  below. Unfortunately
a concise description  of the discrete group
$\Gamma\subset Sp(2r;\IZ)$ does not appear to
be available.

The expression $Z_{\rm Coulomb}$ turns out to
have the
general form:
\eqn\higheruplane{
Z_{\rm Coulomb} = \int_{\CM_{\rm Coulomb} }
[da  d\bar   a]
A(\vec u )^\chi B(\vec u)^\sigma  e^{U + S^2  T_V(\vec u)} \Psi,
}
where $a^I$ are local special coordinates,
$A,B$ are holomorphic automorphic
forms for $\Gamma$ described
below, $U,T$ are   forms associated with
operator insertions,
$\chi,\sigma$ are the Euler character and
signature of $X$, and $\Psi$ is a certain
Narain-Siegel lattice theta function associated to
$H^2(X;\IZ) \otimes \Lambda_{\rm w}(G)$, where
$ \Lambda_{\rm w}(G)$ is the weight lattice of
$G$. The details of this expression
are explained in sections 3 and 4 below.
Some aspects in the derivation of the integrand
were independently worked out in \lns.

It is far from obvious that the integrand of
\higheruplane\ is single-valued on
$\CM_{\rm Coulomb} $. We check this carefully
for the case $G=SU(3)$ and give a less-detailed
general argument for single-valuedness
for $G=SU(N)$ (although we prove the invariance under
the semiclassical monodromies for any simply-laced
group). We do not seriously doubt
that the integrand of \higheruplane\
is single-valued for all $G$ of $r>1$, but our
arguments for this leave  room for
improvement.

The integrand of \higheruplane\ is
singular on $\CD$ and in the weak-coupling
regime at infinity. Hence, some discussion is
required to give rigorous meaning to
the integral \higheruplane.
To do this we need to understand the structure
of the divisor $\CD$ more thoroughly. This divisor
is a stratified space. The maximum dimension stratum
is a union of  several smooth complex codimension one
components $\CD^{(1)}_i$ corresponding physically
to moduli for which a single $u(1)  \subset \liet $ becomes
strong and for which a single monopole hypermultiplet
becomes massless. The strata of
$\CD$   of
higher codimension correspond to singularities in
$\CD$ where  successively larger numbers
of hypermultiplets become massless. We denote the
smooth components of the codimension $\ell$ strata by
$\CD^{(\ell)}_i$. In particular, $\CD^{(r)}_i$ contains
$h(G)$ (the dual Coxeter number)
points corresponding to the supersymmetric
vacua of the $\CN=1$ theory, as well as points corresponding
to multicritical superconformal field theories.

Following the discussion in \mw\ we define the integral
by introducing a cutoff in the weak-coupling regime and
by introducing tubular neighborhoods of  $\CD_i^{(1)}$
and doing a phase integral first over the relevant
special coordinates. The definition of the integration
near $\CD_i^{(\ell)}$ for $\ell>1$ is more problematic
and we only discuss it in full detail in the case
$G=SU(3)$. All this is described in sections 6 and 8.
We expect our  considerations to
generalize to other gauge groups, but again our treatement
leaves room for improvement.

We then implement the wall-crossing argument of
\mw\ by postulating that the metric variation of
$Z_{\rm Coulomb}$ from the singularities near
$\CD_i^{(1)}$ is cancelled by compensating
metric-variation of a mixed Coulomb/monopole theory
which describes the low-energy physics near
$\CD_i^{(1)}$. A consequence of our postulate
\foot{which may be justified physically from
considerations of tunneling between vacua
at finite volume} is that    $Z_{DW}$
must take the form:
\eqn\fullform{
Z_{DW} = Z_{\rm Coulomb} + \sum_i Z_{\CD_i^{(1)}} +
\sum_i Z_{\CD_i^{(2)}}+ \cdots + \sum_i Z_{\CD_i^{(r)}}
}
This is the generalization of equation (1.8) of \mw.
The integrals $Z_{\CD_i^{(1)}} $ along the codimension
one varieties are derived from the wall-crossing of
$Z_{\rm Coulomb} $. This wall-crossing is
described in section 6, and the explicit formulae
for $G=SU(3)$ are derived in  complete detail in equations $(9.1)-(9.6)$ below.
The integrals $Z_{\CD_i^{(1)}}$
themselves have wall-crossing behavior which
is compensated by metric dependence of $Z_{\CD_i^{(2)}}$.
This allows a derivation of the integrand of
$Z_{\CD_i^{(2)}}$ and so on. The procedure terminates
at the codimension $r$ singularities of $\CD$.

The central question of metric variation at a
superconformal point is addressed in section eight.
We analyze the behavior at the Argyres-Douglas
points for $G=SU(3)$ in detail and show that there
cannot be any continuous metric dependence unless
$\sigma<-11$. We also give a general argument that
shows there cannot be any continuous metric
dependence for any signature.
 This argument is based on the blow-up and
wall-crossing formulae. The blow-up formula for the
higher rank case is derived in section seven
by a straightforward
generalization of the derivation in \mw. Using this
formula we can relate the invariants on $X$ to
invariants on a blowdown with sufficiently large
signature that there can be no continuous metric
variation.

In the case $b_{2,+}>1$  only the last
term $\sum_i Z_{\CD_i^{(r)}}$ of \fullform\
is nonvanishing, and indeed only the $\CN=1$
vacua contribute. This allows us to write
the generalization of Witten's formula \monopole\
to $G=SU(N)$ in equation $(9.13)$.

\subsec{Applications  }

Ironically, our work, which was motivated by
topology,  might find its most interesting applications
in physics. In sections 10 and 11 we describe two
applications. In section 10 we use the
behavior of the integrand of \higheruplane\
at superconformal points to deduce
a selection rule for
correlators of $\CN=2, d=4$  superconformal
theories. In section 11 we use the explicit
result $(9.13)$ to study some questions about
the large $N$ behavior of certain correlation functions
in $SU(N)$ SYM theory.

\subsec{Directions for future work}

There are plenty of opportunities for future work.
First, there are technical gaps in our discussion which
have been indicated above. We are confident in our
conclusions, but it should be possible to give a
better treatment of the analysis of $Z_{\rm Coulomb}$.

As in \mw\ the discussion is rather easily extended to
theories with matter. For example, almost the
same formulae hold for $G=SU(N_c)$ with
$N_f$ fundamental hypermultiplets, as long as
the masses of the hypermultiplets are generic.
For special values of the masses some very interesting
things should happen and this remains an interesting
avenue for future research.

Another generalization worth studying is the case of
$SU(N_c)$ with $\CN=4$ supersymmetry perturbed to
$\CN=2$ by the addition of a mass perturbation for
adjoint hypermultiplets. A discussion of $Z_{\rm Coulomb}$
for these theories  is technically challenging but might find interesting
applications in string/M theory.

The generalization of the $u$-plane integrals studied in
\mw\ to the higher rank case is probably only the first
of a series of interesting generalizations of similar integrals
associated to special K\"ahler manifolds.

\newsec{Higher rank $\CN=2$ gauge theories}

In this section, we review some properties of the low-energy
structure of $\CN=2$ gauge theories that we will need in this paper
\swi\klt\dansun. We
then focus on the case of $SU(N)$ Yang-Mills theory, and in particular on the
$SU(3)$ moduli space, which has been explored in some detail \ad\klt. We also
work out some aspects of the solution
near the superconformal, or Argyres-Douglas (AD) points that
will be needed in the rest of the paper.

\subsec{General structure}

The classical moduli space of ${\cal N}=2$ SYM with a rank $r$ gauge
group is determined by the vacuum expectation value of the
field $\phi$, which can always be rotated into the Cartan
subalgebra. Following \klt, we will denote these expectation values
by a vector ${\vec a}$ in the root lattice, and the components
of $\vec a$ , $a^I$ with $I=1, \cdots, r$ will correspond to a basis
of simple roots. The charges will be specified by vectors $\vec q$
expanded in the Dynkin basis ({\it i.e.} the basis of fundamental weights). The
central charges of electric BPS states are then
written as
\eqn\ccharges{
Z_{ \vec q} = {\vec q} \cdot  {\vec a},
}
where the product is given by the usual bilinear form in the
weight lattice. One can then introduce the Casimirs
$u_k$ as Weyl-invariant coordinates
in the classical moduli space.
Singularities in this moduli space are associated semiclassically to
massless gauge bosons, and they occur when $Z_{\vec \alpha}=0$,
where $\vec \alpha$ is a positive root. They are located at the
zeroes of the classical discriminant,
\eqn\clssdisc{
\Delta_0 (\vec u) =\prod_{\vec \alpha >0} Z_{\vec \alpha}^2.
}

The low energy effective action is determined by a prepotential
$\cal F$ which depends on $r$ ${\cal N}=2$ vector multiplets $A^I$. The
VEVs of the scalar components of these vector superfields are
the $a^I$ . The dual
variables and gauge couplings are defined as
\eqn\dualvar{
a_{D, I} = {\partial  {\cal F} \over \partial a^I} ,  \,\,\,\,\,\ \tau_{IJ}=
{\partial ^2 {\cal F} \over \partial a^I \partial a^J}.
}
The moduli space of vacua has a natural
K\"ahler metric given by
\eqn\kalmetric{
(ds)^2 = {\rm Im}  \tau_{IJ} da^I  d \bar a^J,
}
which is invariant under the group ${\rm Sp} (2r, {\IZ})$ (the restriction
to integer valued entries comes from the integrality requirement of the
charges, as we will see in a moment).  The inverse metric will
be denoted by $({\rm Im} \tau)^{IJ}$. Matrices in ${\rm Sp} (2r, {\IZ})$
have the structure
\eqn\sp{
\gamma = \pmatrix{ A & B \cr
C & D},
}
where the $r \times r$ matrices $A$, $B$, $C$, $D$ satisfy:
\eqn\sptwo{
A^t D-C^t B = {\bf 1}, \,\,\,\,\,\ A^tC= C^t A, \,\,\,\,\,\ B^tD= D^t B.
}
The generators of the symplectic group ${\rm Sp} (2r, {\IZ})$ are
\eqn\spgens{
\eqalign{
\CA &= \pmatrix{ A & 0 \cr
0 & (A^t)^{-1} } , \,\,\,\,\,\,\ A \in {\rm Gl}(r, {\IZ}), \cr
T_{\theta} &= \pmatrix{ {\bf 1} & \theta \cr
0 & {\bf 1} } , \,\,\,\,\,\,\  \theta_{IJ} \in {\IZ}, \,\,\,\,\,\,\ \theta^t
= \theta, \cr
\Omega &= \pmatrix{0& {\bf 1} \cr
-{\bf 1} &0},\cr
}
}
The symplectic group acts on the $a^I$, $a_{D,I}$ variables as
$v \rightarrow \gamma v$, where $v^t =(a_{D,I}, a^I)$. In particular, we have
the following transformation properties which will be useful later,
\eqn\spthree{
\eqalign{
{\partial \over \partial a^I} &\rightarrow [(C \tau + D)^{-1}]^J_{~I}
{\partial \over \partial a^J} ,\cr
\tau & \rightarrow (A \tau + B) (C \tau + D)^{-1},\cr
{\rm Im} \tau & \rightarrow [(C {\overline \tau} + D)^{-1}]^t ({\rm Im} \tau)
(C \tau + D)^{-1}.\cr
}
}
\subsec{$SU(N)$}

In the $SU(N)$ case, the quantum theory is described by the hyperelliptic curve
\af \klty:
\eqn\suncurve{
y^2= P(x)^2- \Lambda^{2N}, \,\,\,\,\,\ P(x)= x^N- \sum_{I=2}^{N} u_I x^{N-I},
}
where the $u_I$, $I=2, \dots, N$, are the elementary symmetric polynomials in
the
roots of  $P(x)$.
The quantum discriminant associated to this curve is given by
\eqn\quandelt{
\Delta_\Lambda =
\Lambda^{2 N^2} \Delta_0(u_2, \dots, u_{N-1}, u_N+\Lambda^N) \Delta_0(u_2,
\dots, u_{N-1},
u_N-\Lambda^N)
}
The Coulomb branch of the quantum theory is then given by $\IC^r - \CD$, where
$\CD$ is the vanishing locus of \quandelt. To obtain the couplings $\tau_{IJ}$
and the masses of the BPS states,
one chooses a symplectic homology basis for the genus $r$ Riemann surface
described
by \suncurve, $\alpha_I$, $\beta^I$, $I=1, \cdots, r$, and the basis of
holomorphic
differentials $\omega_I= x^{I-1} dx/y$. The periods of the curve are then
\eqn\periods{
\eqalign{
A_I^{~J}&= \oint_{\alpha_I} \omega_J= {\partial a_{D,I} \over \partial
u_{J+1}}, \cr
B^{IJ}&= \oint_{\beta^I} \omega_J= {\partial a^I \over \partial u_{J+1}},\cr
}
}
where $I$, $J=1, \cdots, r$. The gauge coupling is then given by
\eqn\taueq{
\tau_{IJ}= A_I^{~K} (B^{-1})_{KJ}.
}
One then introduces a meromorphic one form on the
hyperelliptic curve (usually known as Seiberg-Witten differential)
$\lambda_{SW}$
satisfying
\eqn\lsw{
{\partial \lambda_{SW} \over \partial u^{I+1}}= \omega_I,
}
which has the explicit expression \af:
\eqn\lswex{
\lambda_{SW} = { 1 \over 2 \pi i } { \partial P \over \partial x} { x dx \over
y}.
}
The BPS masses are then given by the periods of $\lambda_{SW}$:
\eqn\aad{
a_{D,I}= \oint_{\alpha_I} \lambda_{SW}, \,\,\,\,\  a^I = \oint_{\beta^I}
\lambda_{SW}.
}

Quantum-mechanically, the singularities in the moduli space are associated to
massless
dyons. Their charges will be denoted by $\vec \nu = (\vec g, \vec q)$, where
$\vec g$, $\vec q$ are the $r$-component vectors of magnetic and electric
charges,
respectively. When one of these dyons becomes massless at a certain submanifold
of the
moduli space, one of the cycles of the hyperelliptic curve degenerates and
there is an
associated monodromy given by
\eqn\monod{
M_{\vec \nu} = \pmatrix{ {\bf 1} + {\vec q} \otimes {\vec g} & {\vec q} \otimes
{\vec q}\cr
- {\vec g} \otimes {\vec g} & {\bf 1}- {\vec g} \otimes {\vec q} \cr}.
}

\subsec{$SU(3)$ and the AD points}

In the case of ${\cal N}=2$ SYM theory with gauge group $SU(3)$, the moduli
space is
parametrized
by the Casimirs $u=u_2$, $v=u_3$. There is a discrete, anomaly-free subgroup
${\IZ}_6$
of the $R$-symmetry which acts as $u \rightarrow {\rm e}^{2 \pi i /3} u$, $v
\rightarrow -v$. The quantum discriminant
is given by
\eqn\suthreedisc{
\Delta_{\Lambda}=\Lambda^{18} [ 4 u^3 -27 (v+ \Lambda^3)^2] [ 4 u^3 -27 (v-
\Lambda^3)^2].
}
There are two codimension one submanifolds given by $\Delta_0 (u, v \pm
\Lambda^3)=0$, which
intersect in the three ${\IZ}_2$ vacua $4u^3= (3 \Lambda^2)^3$, $v=0$.
At these points there are
two mutually local dyons becoming massless, and when ${\cal N}=2$ is softly
broken
down to
${\cal N}=1$ with a superpotential $\Tr \Phi^2$, they give the three vacua of
${\cal N}=1$
SYM \af. The charges $(n_m^1, n_m^2; n_e^1, n_e^2) $
of these states are the following \klt:
\eqn\strongspec{
\eqalign{
\vec \nu_1&= (1,0; 0,0), \,\,\,\,\,\  \vec \nu_2=(0,1;0,0), \cr
\vec \nu_3 &=(0,1; -1,2),  \,\,\,\,\,\  \vec\nu_4=(-1,-1; 2,-1), \cr
\vec \nu_5 &=(-1,-1; 1,-2),  \,\,\,\,\,\  \vec \nu_6=(1,0; 2,-1). \cr
}
}
The charges
in the same row in \strongspec\ are mutually local. The first row
corresponds to the ${\IZ}_2$ vacuum at $u_{1} =(27/ 4)^{1/3} \Lambda^2$,
$v=0$.  The second and third rows correspond to the vacua located at
$u_2={\rm e}^{2\pi i /3}u_1$, $v=0$, and $u_3={\rm e}^{4\pi i /3}u_1$, $v=0$,
respectively.
In fact one can find a matrix $U$ \klt\ which implements the ${\IZ}_3$
symmetry in moduli space,
\eqn\umatrix{
U=\pmatrix{-1 & -1& 1&2 \cr
1&0&-2& 1\cr
0&0&0&-1\cr
0&0&1&-1\cr
}.
}
One can check that, acting on the right on $\vec \nu_1$, $\vec\nu_2$, we obtain
the other two pairs of massless states at the ${\cal N}=1$ points, {\it i.e.},
$\vec
\nu_1 U^{-1}= \vec \nu_3$, $\vec \nu_2 U^{-1}= \vec \nu_4$, etc.

There are also singular points on each of the submanifolds (also called
${\IZ}_3$
vacua) at the points $u=0$, $v = \pm \Lambda^3$. These are the Argyres-Douglas
(AD) points, where
three mutually non-local hypermultiplets become massless \ad. We will be
particularly interested
in the behaviour of the theory near these points. Let's focus on the point $v=
\Lambda^3$, $u=0$
(the behaviour near the other AD point can be obtained using the ${\IZ}_6$
symmetry which sends
$v \rightarrow -v$). The states that become massless near this point are $\vec
\nu_ 1$, $\vec \nu_3$ and
$\vec \nu_5$. The symplectic transformation $T_{\theta} \Omega^{-1} {\cal A}$,
where
\eqn\symplead{
A= \pmatrix{ -1&0 \cr
    2& 1\cr }, \,\,\,\,\,\ \theta=\pmatrix{ -1 & -1 \cr
                                                                    -1&
\theta_{22}\cr}, \,\,\ \theta_{22} \in {\IZ},
}
gives a basis where all the states are charged only with respect to the first
$U(1)$ factor. Applying ${\cal A}^{-1} \Omega T_{\theta}^{-1}$ on the right to
the charge vectors $\nu_i$, we find that the new
charges $\vec \nu \cdot {\cal A}^{-1}
\Omega T_{\theta}^{-1}$
are (in this new basis) $(n_e, n_m) = (-1,0)$
for $\vec \nu_1$, $(1,-1)$ for $\vec \nu_3$, and $(0,1)$ for $\vec \nu_5$,
{\it i.e.} we have one electron, one dyon, and one monopole. In this basis, the
hyperelliptic
curve degenerates and at leading order in $u$, $v-\Lambda^3$, it splits into a
``small" torus whose periods go to zero (and correspond
to $a^1$, $a_{D,1}$) and a ``large" torus whose periods $a^2$, $a_{D,2}$ are of
order $\Lambda$.

We introduce now the useful parameters $\epsilon$, $\rho$ around the AD point,
defined by
\eqn\adpar{
u= 3 \epsilon^2 \rho, \,\,\,\,\  v-\Lambda^3 = 2 \epsilon^3.
}
The variable $\rho$ parametrizes the direction along which we approach the AD
point
in the $u$, $v$ moduli space. The equation defining the small torus near the AD
point is
given by
\eqn\rhocurve{
w^2=z^3-3 \rho z -2,
}
with discriminant
\eqn\rhodisc{
\Delta_{\rho}= 4 \cdot 27 (\rho^3-1),
}
and the Seiberg-Witten differential on the curve \rhocurve\ degenerates to
\eqn\swdiffad{
\lambda_{SW} = p {\epsilon^{5/2} \over \Lambda^{3/2} } w dz,
}
where $p$ is some constant that depends on the normalization of $\lambda_{SW}$.
The small
torus theory gives us the dependence on $\rho$ for the leading terms in
$\epsilon$ of $a^1$,
$a_{D,1}$.  We can put the curve \rhocurve\ in Weierstrass form and compute
$a^1$, $a_{D,1}$
explicitly (at leading order in $\epsilon$) in terms of the periods of the
curve $\omega_{\rho}$,
$\omega_{\rho,D}$ (with ${\rm Im}(\omega_{\rho,D}/\omega_{\rho})>0$) :
\eqn\asad{
a^1= {\epsilon^{5/2} \over \Lambda^{3/2} } f(\rho), \,\,\,\,\  a_{D,1}=
{\epsilon^{5/2} \over \Lambda^{3/2} } f_D(\rho),
}
where
\eqn\efs{
f(\rho)= {48 p \over 5} (\rho \eta -
{\omega_{\rho}\over 8}), \,\,\,\,\  f_D(\rho)= {48 p \over 5} (\rho \eta_D -
{\omega_{\rho,D}\over 8})
}
In this equation, $\eta= \zeta(\omega_{\rho}/2)$,
$\eta_D=\zeta(\omega_{\rho,D}/2)$ are the
usual values of the Weierstrass zeta function at the half-periods.

The curve \rhocurve\ has singularities when $\rho^3=1$ and also at infinity. At
$\rho^3=1$
we have $A_0$ singularities and the behaviour of $\tau (\rho)$ is
\eqn\azero{
\tau({\rho})= {1 \over 2 \pi i} {\rm log} (\rho-\rho_k),
}
where $\rho_k={\rm e}^{2\pi i k/3}$, $k=0,1,2$ are the corresponding
singularities. At $\rho
\rightarrow \infty$, there is an $H_1$ singularity (Kodaira's type III) and the
monodromy is given
by $S^{-1}$.
The behaviour of $\tau (\rho)$ is given by
\eqn\tauinfty{
\tau(\rho) = i + { C \over \rho^{3/2}} + \cdots,
}
where $C$ is a nonzero constant.

The behaviour of  $a^2$, $a_{D,2}$ (the ``long periods") can be found \ad \ms\
to be
\eqn\atad{
a^2= b \Lambda + c { u \over \Lambda} + d { v-\Lambda^3 \over \Lambda^2}
+\cdots, \,\,\,\,\ a_{D,2}= b_D \Lambda
+ c_D {u \over
\Lambda} + d_D { v-\Lambda^3 \over \Lambda^2}+\cdots,
}
where $b$, $c$, $d$, $b_D$, $c_D$ and $d_D$ are non-zero constants. From these
explicit
expressions we can compute the matrix of periods of the hyperelliptic curve,
$B^{IJ}$, at leading order:
\eqn\adperiods{
\pmatrix{ {\partial a^1 \over \partial u} & {\partial a^1 \over \partial v} \cr
{\partial a^2 \over \partial u} & {\partial a^2 \over \partial v}\cr} =
\pmatrix{ { \epsilon^{1/2} \over 3 \Lambda^{3/2}} f'(\rho) & {\epsilon^{-1/2}
\over 3 \Lambda^{3/2}} H(\rho) \cr
{c \over \Lambda} & {d \over \Lambda^2} \cr},
}
where the derivatives are with respect to $\rho$ and
\eqn\period{
f'(\rho) = 12 p \eta, \,\,\,\,\ H(\rho)= {5 \over 4} f(\rho)- \rho f'(\rho)=
-{3 \over 2} p \omega_{\rho}.
}
The matrix $A_I^{~J}$ has a similar expression in terms of $c_D$, $d_D$ and
$f_D(\rho)$.
We then find that
\eqn\detb{
\det {\partial u_{I+1} \over \partial a^J} = {2 \Lambda^{5/2} \over p c} {
\epsilon^{1/2} \over \omega_{\rho}} + {\cal O}(\epsilon^{3/2}).
}
The gauge couplings can be also computed in a straightforward way and
are given by
\eqn\adtaus{
\eqalign{
\tau_{11} = & \tau(\rho) - {\epsilon d  \over c \Lambda} { f'_D (\rho) \over
H(\rho)}+ {\cal O}(\epsilon ^2) ,\cr
\tau_{12} = & -{4\pi i p \over c \Lambda^{1/2}} {\epsilon^{1/2}\over
\omega_{\rho}}+ {\cal O}(\epsilon^{3/2}), \cr
\tau_{22}=& {c_D \over c} - {\epsilon d_D   \over c \Lambda} { f' (\rho) \over
H(\rho)} +{\cal O}(\epsilon^{2}),\cr}
}
and $c_D/c= {\rm e}^{ \pi i /3}$ is the period of the large torus at the AD
point (some aspects of
the behaviour of the couplings at the AD point have been investigated in \ky).

Finally, we will need  the behaviour of the third derivatives of the
prepotential
near this point (and in particular their leading behaviour in $\epsilon$).
These are given by:
\eqn\prepotad{
\eqalign{
\CF_{111}=& - { \Lambda^{3/2} \epsilon^{-5/2} \over H(\rho)} \rho {d \tau(\rho)
\over d \rho} + {\cal O}(\epsilon^{-3/2}), \cr
\CF_{112}=&  { 5 \Lambda  \epsilon^{-2} \over 12 c } {f (\rho) \over H(\rho)}
{d \tau(\rho) \over d \rho} +{\cal O}(\epsilon^{-1}), \cr
\CF_{122}=& { \Lambda^{1/2} \epsilon^{-3/2} d_D \over c H(\rho)} \bigg[ \Bigl(
{ f' (\rho)
\over H(\rho) } \Bigr)' - {1 \over 2} { f' (\rho)
\over H(\rho) } \biggr] +{\cal O}(\epsilon^{-1/2}), \cr
\CF_{222}= &{ \epsilon^{-1} d_D \over 6 c^2 } \bigg[ \Bigl( { f' (\rho)
\over H(\rho) } \Bigr)^2 - {5 \over 2} { f (\rho)
\over H(\rho) }  \Bigl( { f' (\rho)
\over H(\rho) } \Bigr)' \biggr]+ {\cal O}(1 ).\cr
}
}
This behaviour is consistent with the $R$-charge assignment near the
superconformal point,
$R(a^1)=1$, $R(a^2)=R(u)=4/5$, $R(\CF)=2$.

\newsec{The twisted effective theory on the Coulomb branch}

To study the twisted supersymmetric ${\cal N}=2$ SYM theory on a four-manifold
$X$, we
consider the
low-energy description encoded in the solution presented in the last section.
The
procedure we will follow is a straightforward generalization of the one
presented
in \mw. The field content of the low-energy theory consists of  $r$ twisted
abelian
${\cal N}=2$ vector multiplets. The $\CQ$-transformations are given by
\eqn\toptmns{
\eqalign{
[\overline{\CQ}, A^I ]  =  \psi^I ,
\quad & \quad
[\overline{\CQ}, \psi^I]  =  4 \sqrt{2} d a^I,\cr
[\overline{\CQ}, a^I]  = 0, \quad & \quad
[\overline{\CQ}, \bar a^I]  =    \sqrt{2}i \eta^I,\cr
[\overline{\CQ}, \eta^I]  = 0 , \quad & \quad
[\overline{\CQ}, \chi^I]  = i( F_+^I  - D_+^I),  \cr
[\overline{\CQ}, D^I]  =   (d \psi^I)_+.  \cr }
}
We will also need the action of the one-form operator $G$, which gives
a canonical solution to the descent equations (this operator was denoted
by $K$ in \mw). It is given by
\eqn\actkay{
\eqalign{
[G , a^I]  = {1 \over 4 \sqrt{2}} \psi^I ,\quad & \quad
[G, \psi^I]  = - 2(F^I_-+ D^I), \cr
[G,  A^I ] = - 2i \chi^I,\quad & \quad
[G, \bar a^I]  = 0, \cr
[G, \eta^I]  = -{i \sqrt{2} \over 2} d \bar a^I , \quad & \quad
[G, D^I]  = -{3i \over 4} * d \eta^I + {3 i \over 2} d \chi,\cr
[G, \chi^I]  = -{3i \sqrt{2} \over 4} * d \bar a^I.\cr}
}

The twisted effective Lagrangian can be written in a
manifestly topological way as:
\eqn\manfsttop{
\eqalign{
&{i \over 6 \pi} G^4 \CF(a^I) + { 1 \over 16 \pi} \{ \overline  \CQ, \overline
\CF_{IJ}
\chi^I (D+ F_+)^J\} - { i \sqrt{2} \over 32 \pi}
\{ \overline  \CQ, \overline \CF_{I }  d * \psi^I \}\cr
& - {{\sqrt 2}i \over 3 \cdot 2^5 \pi} \{ \overline \CQ, \overline \CF_{IJK }
\chi_{\mu \nu}^I
\chi^{\nu \lambda J} \chi_{\lambda}^ {~ \mu K} \},\cr}
}
which may be expanded out to give:
\eqn\chcklag{
\eqalign{
& { i \over  16 \pi}   \bigl( \overline \tau_{IJ}  F_+^I  \wedge F_+^J
+   \tau_{IJ} F_-^I \wedge F_-^J \bigr)
 +  {1 \over  2 \pi}    (Im \tau_{IJ})  da^I \wedge * d\bar a^J
     - {1 \over  8 \pi}   (Im \tau_{IJ}) D^I \wedge *D^J \cr
&- {1 \over  16 \pi}  \tau_{IJ}  \psi^I \wedge * d \eta^J
+ {1 \over  16 \pi} \overline  \tau_{IJ} \eta^I \wedge d * \psi^J
+ {1 \over  8 \pi}  \tau_{IJ} \psi^I \wedge d \chi^J
- {1 \over  8 \pi} \overline  \tau_{IJ} \chi^I \wedge (d \psi^J)_+
\cr
&+ {i \sqrt{2}   \over  16 \pi } \overline \CF_{IJK} \eta^I \chi^J
\wedge (D_+ + F_+ )^K
 - {i \sqrt{2}   \over  2^7 \pi } \CF_{IJK}
(\psi^I \wedge \psi^J) \wedge  ( F_-  +  D_+)^K
\cr
 & + {i \over  3 \cdot 2^{11}  \pi  }    \CF_{IJKL}
 \psi^I \wedge \psi^J \wedge \psi^K \wedge \psi^L   - {{\sqrt 2}i \over 3 \cdot
2^5 \pi}
\{ \overline \CQ, \overline \CF_{IJK }  \chi_{\mu \nu}^I
\chi^{\nu \lambda J} \chi_{\lambda}^{~ \mu K} \} .\cr }
}

It is important to notice that the part of the action involving the fourth
descendant of the prepotential can be written as
a ${\overline \CQ}$-exact term plus terms which are topological ({\it i.e.}
they do not
involve the
metric of the four-manifold $X$):
\eqn\qexact{
\eqalign{
&{i \over 6 \pi} G^4 \CF(a^I)= \cr
& \,\,\,\ \biggl\{ {\overline Q}, \tau_{IJ} \Bigl[
-{{\sqrt 2} i \over 2^5\cdot \pi}
\psi ^I \wedge *d{\bar a}^J -{1 \over 16 \pi} \chi ^I \wedge(F^-+ D)^J \Bigr] +
{{\sqrt 2} \over 2^7 \cdot \pi} {\CF}_{IJK} \psi^I \wedge \psi^J \wedge \chi^K
\biggr\}\cr
&\,\,\,\ + { i \tau^{IJ} \over 16 \pi} F^I \wedge  F^J -{i \sqrt{2}   \over
2^7 \pi }
\CF_{IJK}
(\psi^I \wedge \psi^J) \wedge  F^K + {i \over  3 \cdot 2^{11}  \pi  }
\CF_{IJKL}
 \psi^I \wedge \psi^J \wedge \psi^K \wedge \psi^L,\cr}
}
where integration over $X$ is understood.

\newsec{The integrand in the higher rank case}

The $u$-plane integral in the higher rank case is
given by a general expression of the form:
\eqn\higheruplane{
Z_u(p,S;m_i,\tau_0) = \int_{\CM_{\rm Coulomb} }
[da  d\bar   a]
A(\vec u )^\chi B(\vec u)^\sigma  e^{U + S^2  T_V(\vec u)} \Psi,
}
where $\Psi$ is a certain lattice theta function. We will explain in some
detail the structure of the different terms involved in
\higheruplane.  The resulting expression, as we will see, holds for any
simply-laced gauge
group.

\subsec{ The observables}

In \higheruplane\ the $0$-observable is a general invariant function
$U$ on the Lie algebra. This generalizes $2p u$ in the
rank one case. We will restrict
attention to expressions linear in the Casimirs of the group,
\eqn\zerobs{
U = \sum_{I=2}^{r+1}  p^I {\rm Tr} \phi^I.
}
Here we are using the standard notation for the Casimirs of $SU(N)$. The VEVs
of these operators can be related to the symmetric polynomials in
\suncurve\ which
parametrize the quantum moduli space by standard results on symmetric
functions.

The $2$-observable is obtained by canonical descent from another
general function $V$.  When $2$-observables are included one has to take into
account
contact terms, denoted by $T_V(\vec u)$. We will discuss them below. For
simplicity,
we will restrict attention to 2-observables obtained
from the quadratic Casimir,  $V=u_2$. Other 2-observables involve new contact
terms discussed in \lns. In general, the 2-observable is obtained
as in \mw\ using the one-form operator $G$, which gives canonical solutions to
the descent equations. In the rank $r$ case we have
\eqn\obsvs{
G^2 u_{I}  = { 1 \over 32} {\p ^2 u_{I}  \over  \p a^J \p a^K } \psi^J \wedge
\psi^K -
{ \sqrt{2}\over  4} {\p u_{I} \over  \p a^J} (F_-  +   D_+)^J.
}
The two-observable associated with a surface $S$ is given by
\eqn\twoobs{
{\tilde I}(S)= { i \over \pi {\sqrt 2}} \int_{S} G^2 u_{I},
}
where we use the normalization of \mw.

The four-observables come again from a general function $W$. Using the
canonical solution
to the descent equations we see that they merely shift $\tau_{IJ} \rightarrow
\tau_{IJ} +
W_{IJ}$. This will involve further contact
terms, which can be written by a process of covariantizing
derivatives.

\subsec{The measure factor}

The $A,B$ functions in \higheruplane\  are the higher rank generalization of
the
gravitational factors considered in \witteni\mw. They are given by:
\eqn\afunction{
A^\chi = \alpha^\chi \biggl( \det {\p u_I \over  \p a^J} \biggr)^{\chi/2}
}

\eqn\bfunction{
B^\sigma = \beta^\sigma  \Delta_\Lambda^{\sigma/8}
}

This may be proved by a modification of the argument
of  \witteni\mw.  The twisted theory with
gauge group $G$ has a gravitational contribution to the
anomaly given by $-({\rm dim} G)(\chi + \sigma)/2$. In the semiclassical regime
the effective $U(1)^r$ theory gives the anomaly $-r(\chi + \sigma)/2$. The
remaining anomaly
should be carried by the measure factor in the semiclassical region. On the
other hand,
near the divisor where a single hypermultiplet becomes massless, there is an
accidental
low-energy $R$-symmetry given by $-\sigma/4$ which should also show up
in the measure factor in this region.

We first check that the $B^\sigma$ factor gives the needed behaviour
for the $\sigma$ dependence. Near the divisor where a single hypermultiplet
becomes massless, the quantum discriminant has the structure
\eqn\strongdisc{
\Delta_{\Lambda}^{ \sigma /8} \sim Z^{\sigma /8} \tilde \Delta_{\Lambda},
}
where $Z$ is the transverse coordinate at the divisor and $\tilde
\Delta_{\Lambda}\not=
0$ along it. As $Z$ has $R$-charge two, we see that  $B^\sigma$ gives the
right behaviour. On the other hand, in the semiclassical region we have that
\eqn\semidisc{
\Delta_{\Lambda} \sim (\Delta_{0})^2.
}
The $Z_{\alpha}$ have $R$-charge $2$. As there are $({\rm dim} G-r)/2$ positive
roots, the
$R$-charge of $\Delta_{\Lambda}$ in the semiclassical region is given by
$4({\rm dim} G -r)$, and
again we find the right charge. As $\Delta_{\Lambda}$ is a modular form of
weight zero,
$B/\Delta_{\Lambda}$ has no zeroes and is a constant. This proves  \bfunction.

We now consider the $A^{\chi}$ factor. The $R$-charge at the semiclassical
regime
is easily computed to give $\chi ({\rm dim} G -r)/2$, again in agreement with
the
behaviour we need. On the other hand, we have to check that (in the appropriate
local variables) this factor does not have zeros or singularities on the moduli
space.
Notice that, at a generic point in the moduli space, the $A^{\chi}$ factor can
be written
as
\eqn\afactor{
A^{\chi } = \alpha^{\chi}  \bigl( {\rm det} B^{IJ} \bigr)^{-\chi/2},
}
where ${\rm det} B^{IJ}$ is the first minor of the period matrix of the
hyperellyptic curve \suncurve, and is given in \periods. It follows
from the Riemann bilinear relations that this minor is nonsingular. On a
divisor where a
hypermultiplet
becomes massless, there are good coordinates $a^I$ around it in the sense that
the Jacobian
of the change of variables from $u_I$ to $a^J$ is nonsingular, and again we see
that in the
appropriate variables this factor has no zeros or singularities.
Since   $\det {\p
u_{I+1} /  \p a^J} $
is a modular form of weight $(-1,0)$, we have proved \afunction. We will
comment on the
constant $\alpha$ below.

\subsec{The lattice $\Gamma$ and generalized Stiefel-Whitney classes}

The function $\Psi$ in \higheruplane, as we will see, involves the evaluation
of the photon
partition function for the effective $U(1)^r$ theory. Therefore, it includes
a sum over electric line bundles \witteni. We will consider theories with a
non-abelian magnetic flux. This is possible, for instance, in the case of an
$SU(N)$ theory, because the gauge group is actually $SU(N)/{\IZ}_N$ (provided
all
fields are in the adjoint representation of the group).
A bundle $E$ with this gauge group is characterized up to isomorphism by two
topological
invariants : the instanton number and the generalized Stiefel-Whitney class (or
non-abelian
magnetic flux) ${\vec w}_2 (E)
\in H^2(X, {\IZ}_N)$. For a gauge group $G$, the non-abelian magnetic flux
${\vec w}_2 (E)$ takes values in $H^2(X, \pi_1 (G))$. Equivalently \vw, for any
simply-laced gauge group, the
magnetic
fluxes are cohomology classes in $H^2(X, \Lambda_{\rm w} /
\Lambda_{\rm r})$, where $\Lambda_{{\rm w} ({\rm r})}$ are the weight and root
lattices
of the group, respectively.  For every weight lattice, there is a set of
weights called minimal
weights which are in one-to-one correspondence with the cosets $\Lambda_{\rm w}
/
\Lambda_{\rm r}$ (\humphreys, p. 72). There are in general
$c-1$ minimal weights, where $c=\det C$ is the ``index of connection", that is,
the determinant of the Cartan matrix
(notice that
$c$ is precisely the order of  $\Lambda_{\rm r}$ in $\Lambda_{\rm w}$). The set
of minimal
weights is in general a subset of the set of fundamental weights. We will
denote these weights by ${\vec m}_I$, $I=1, \cdots, c-1$. In the case of
$SU(N)$, they are just the
fundamental weights $ \vec w_I$, $I=1, \cdots, N-1$. The electric line bundles
are then classified by vectors of the form:
\eqn\eleclinebdle{
\vec \lambda ={\vec \lambda}_{\IZ} + \vec v, \,\,\,\,\   {\vec
\lambda}_{\IZ}=\sum_{I=1}^r \lambda^I_{\IZ}   \vec \alpha_I , \,\,\,\,\,\ \vec
v= \sum_{I=1}^{c-1}
\pi^I  \vec m_I,
}
where $\vec \alpha_I$ is a set of simple roots. In this expression,
$\lambda^I_{\IZ}, \pi^I$
are all integer classes in $H^2(X;\IZ)$.
The $\pi^I$ are fixed, and represent a choice of
$\vec w_2(E)\in H^2(X, \Lambda_{\rm w} /
\Lambda_{\rm r})$ lifted to $H^2(X, \Lambda_{\rm r})$. Notice that we can
always expand the minimal weights in the basis of simple roots:
\eqn\minimal{
{\vec m}_I = \sum_{J=1}^r m_I^{~J} {\vec \alpha}_J, \,\,\,\,\ I=1, \cdots, c-1,
\,\,\,\,\
 m_I^{~J}  \in {1 \over c} {\IZ},
}
therefore we can write
\eqn\eleclinebdlep{
\vec \lambda = \sum_{I=1}^{r} \lambda^I {\vec \alpha}_I, \,\,\,\,\
\lambda^I =\lambda^I_{\IZ} + \sum_{J=1}^{c-1} m_J^{~I } \pi^J \in {1 \over c}
H^2(X, {\IZ}).}
For $SU(N)$, one has $m_I^{~J}=(C^{-1})_I^{~J}$, where $C_I^{~J}$ is the Cartan
matrix. Finally, later we will need the result that the instanton number of the
original bundle $E$
satisfies \vw\
\eqn\instanton{
c_2(E) = -{{\vec v} \cdot {\vec v} \over 2} \,\ {\rm mod}~1.
}

\subsec{The lattice sum}

The lattice sum $\Psi$ appearing in \higheruplane\ is obtained after
integrating over
the zero modes of the fields, integrating out the auxiliary fields (after
including the
$2$-observable \twoobs) and taking into account the photon partition function.
The procedure is entirely analogous to the one presented in \mw\ for $SU(2)$.
The only difference
is that the photon partition function includes now a factor $(\det {\rm Im}
\tau)^{-1/2}$ (for
simply-connected manifolds). We also have $r$ zero modes for $\eta^I$ as well
as for
$\chi^I$ (when $b_2^+=1$). Because of the argument based on the scaling of the
metric of \mw, the contribution of the Coulomb branch vanishes if $b_2^+>1$. We
also
write $F^I= 4 \pi \lambda^I$, which is the appropriate normalization for
the line bundles involved in the sum. The lattice $\Gamma$ of $\lambda^I$ has
been already specified
for the general case in which we have non-abelian magnetic fluxes. After
taking all this into account, we finally obtain a formula for the factor $\Psi$
in \higheruplane:
\eqn\hirkthet{
\eqalign{
\Psi& =(\det {\rm Im} \tau)^{-1/2}
 \exp\bigl[   { 1 \over  8 \pi  }V_J ({\rm Im} \tau)^{JK} V_K  S_+^2 \bigr]
\sum_{\lambda \in \Gamma }   \cr
& \exp\biggl[ - i \pi {\overline \tau}_{IJ} (\lambda_+^I,\lambda_+^J)
- i \pi   \tau_{IJ} (\lambda_-^I,\lambda_-^J)
- i \pi ((\vec \lambda -\vec \lambda_0) \cdot  \vec \rho,  w_2 (X))  - i   V_I
(S,\lambda_-^I) \biggr] \cr
& \int \prod_{I=1}^r d \eta^I d \chi^I
\exp\biggl\{ -{i  \sqrt {2} \over 16 \pi} {\overline {\cal F} }_{IJK}\eta^I
\chi^J [4\pi (\lambda_{+}^K, \omega)
+ i ({\rm Im} \tau)^ {KL} V_L (S, \omega) ]
 \cr
& + {1 \over 64 \pi } {\overline {\cal F}}_{KLI} ({\rm Im}  \tau)^{IJ}
{\overline {\cal F} }_{JPQ}
\eta^K  \chi^L \eta^P \chi^Q  \biggr\} \cr}
}
Here $V_I = {\p V \over  \p a^I}$. The phase factor $\exp[- i \pi ( (\vec
\lambda
-\vec \lambda_0) \cdot  \vec \rho,  w_2 (X))]$ can be derived by a
generalization of Witten's analysis in
\witteni\ (see \lns\ for a derivation along these lines). We found it
(independently) from
invariance of the Coulomb integral under the semiclassical  monodromy. $\vec
\lambda_0$ is an element in $\Gamma$ such that $\vec \lambda -\vec
\lambda_0 \in
H^2(X, \Lambda_{\rm r})$, and corresponds to a choice of
orientation of the higher rank instanton moduli spaces. Notice that its
inclusion is
necessary in order for the phase factor to be defined independently of the
integral lift we choose for $\vec \lambda$. One should also
include in the lattice sum a global phase factor depending on the generalized
Stiefel-Whitney class $\vec v$, in order to obtain invariants that are real. In
the
$SU(2)$ case, this factor turns out to be ${\rm e}^{i \pi \vec v \cdot \vec v}
=
{\rm e}^{i\pi w_2(E)^2/2}$ \mw. We will find the appropriate factor for $SU(N)$
after computing the resulting invariants in section 9.

In the rank one case, this lattice sum is
related to the sum $\Psi_{r=1}$ introduced in \mw\ as follows:
\eqn\relrkone{
\eqalign{
\Psi= &-{i {\sqrt 2} \over 4} {1 \over y^{1/2}} {d \overline \tau \over
d\overline a}
\sum_{\lambda \in \Gamma} (-1)^{(\lambda - \lambda_0) \cdot w_2(X)}   \biggl[
(\lambda,
\omega) + {i \over 4 \pi y} {du \over da} (S, \omega) \biggr] \cr
& \cdot
\exp \biggl[ -i \pi {\overline \tau} (\lambda_+)^2 - i \pi  \tau (\lambda_-)^2
-i {du \over da} (S, \lambda_-) \biggr] \cr
=&-{i {\sqrt 2} \over 4}{1 \over y^{1/2}} {d \overline \tau \over d\overline a}
 \exp \bigl[ -{1 \over 8 \pi y} \bigl( {du \over da}\bigr)^2 S^2 \bigr]
\Psi_{r=1} ,
\cr}
}
where $\tau=x+iy$.

We can explicitly evaluate the
Grassmann integral in the rank two case, with the result:
\eqn\rktwoint{
\eqalign{
&\int \prod_{I=1,2} d \eta^I d \chi^I
\exp\biggl\{ -{i  \sqrt {2} \over 16 \pi} {\overline {\cal F} }_{IJK}\eta^I
\chi^J [4\pi (\lambda_{+}^K, \omega)
+ i ({\rm Im} \tau)^{KL} V_L (S, \omega)]
 \cr
& + {1 \over 64 \pi } {\overline {\cal F}}_{KLI} ({\rm Im}\tau_{IJ})^{-1}
{\overline {\cal F} }_{JPQ}
\eta^K  \chi^L \eta^P \chi^Q  \biggr\} \cr
&=-{1\over \pi ^2  2^7} \bigl({\overline {\cal F} }_{11I} {\overline {\cal F}
}_{22J} -
{\overline {\cal F} }_{12I} {\overline {\cal F} }_{12J} \bigr) \biggl\{
-4 \pi  ({\rm Im} \tau)^{IJ} \cr
&+[4\pi (\lambda_{+}^I, \omega)
+ i ({\rm Im} \tau)^{IK} V_K (S, \omega) ]  [4\pi (\lambda_{+}^J, \omega)
+ i ({\rm Im} \tau)^{JL} V_L (S, \omega)]  \biggr\} \cr}
}

In general, the integration over the Grassmann variables will give a factor of
the
form $\det _{IJ} ({\overline \CF}_{IJK} \lambda_+^K) + \cdots$, where the
remaining terms
should be regarded as contact terms.

A more compact expression for \hirkthet\ can be
found if we introduce $r$ bosonic auxiliary variables $b^I$:
\eqn\findiml{
\eqalign{
\Psi = \sum_{\lambda \in \Gamma}
&
\int \prod_{I=1}^r d\eta^I d \chi^I \int_{-\infty}^{+\infty} \prod_{I=1}^r
d b^I
\exp\biggl[ - i \pi {\overline \tau}_{IJ} (\lambda_+^I , \lambda_+^J)  -
i \pi \tau_{IJ} (\lambda_-^I , \lambda_-^J)
\cr
&+ {1 \over  8 \pi} b^I ({\rm Im } \tau)_{IJ} b^J - i V_I (S, \lambda_-^I) -{i
\over 4\pi} V_I
(S, \omega) b^I \cr
& - { i \sqrt{2} \over  16 \pi} {\overline \CF}_{IJK} \eta^I \chi^J (b^K + 4
\pi
\lambda_+^K)  - i \pi ((\vec \lambda -{\vec \lambda}_0) \cdot \vec \rho ,  w_2
(X)) \biggr] . \cr}
}
We emphasize that the integral in \findiml\ is finite -dimensional, and not a
path integral.
This expression can be formally considered as the partition function
of a finite-dimensional topological ``field" theory, obtained from the
original one after restriction to the sector of harmonic forms.
The topological invariance is obtained from \toptmns\ and reads:
\eqn\toptfin{
\eqalign{
[\overline{\CQ}, \lambda ] =  0,
\quad & \quad
[\overline{\CQ}, a^I]   = 0,\cr
[\overline{\CQ}, \bar a^I] =   \sqrt{2}i \eta^I,
\quad & \quad
[\overline{\CQ}, b^I]  =  0,\cr
[\overline{\CQ}, \eta^I ] = 0,
\quad & \quad
[\overline{\CQ}, \chi^I ]   = i(4 \pi  \lambda_+^I  - b^I) .\cr
}
}
We can consider minus the exponent in \findiml\
as the (euclidean) action $S_E$ of this topological field theory.
It is $\overline {\CQ}$-closed.

\subsec{The contact term}

As explained in \mw, when $2$-observables are taken into account  there are
possible contact terms in the low-energy description. As will become clear
in the next section, the contact term
$T_V$ must be such that
\eqn\dualct{
{\widehat T}_V (\vec u) = T_V (\vec u) + {1 \over 8 \pi} V_J ({\rm Im}
\tau)^{JK} V_K
}
is duality invariant.  We give its form for $V=u_2$ and
a general $SU(N)$ theory with $N_f$ matter hypermultiplets, $N_f \le 2N$,
following the approach of \mw \mm. Introduce the parameter
$\tau_0$ as $\Lambda_{N,N_f}^{2N-N_f}= {\rm e}^{i \pi \tau_0}$ for the
asymptotically
free theories, and as the microscopic gauge coupling for the theories with
$N_f=2N$. The
prepotential verifies the relation \matone \ey \sty
\eqn\matgeneral{
{\partial \CF \over \partial \tau_0} = { 1 \over 4} u_2,
}
and under a symplectic transformation we have the following behaviour,
\eqn\symtt{
{\partial ^2 \CF \over \partial \tau_0^2} \rightarrow
{\partial ^2 \CF \over \partial \tau_0^2}-{\partial^2 \CF \over \partial \tau_0
\partial a^I}
[(C \tau + D)^{-1}]^I_{~J} C^{JK} {\partial^2 \CF \over \partial \tau_0
\partial a^K}.
}
If we take into account that $V_I=4 ({ \partial^2  \CF /\partial a^I \partial
\tau_0})$, we see that
the shift of \symtt\ has the same structure as the shift of the second term
in \dualct\ under symplectic transformations.
It follows that the contact term can be written as
\eqn\contactp{
T(\vec u)  = { 4 \over \pi i}  {\p^2 \CF \over  \p \tau_0^2},
}
In some cases we can use the homogeneity properties of $u_2$ to write more
explicit
expressions for $T(\vec u)$. In the case of $N_f < 2N$ massless hypermultiplets
we have
\eqn\contiii{
T(\vec u)  ={1 \over 2N-N_f}  \Bigl( 2 u_2 - \sum_I a^I {\p u_2 \over  \p
a^I}\Bigr).
}
Notice from this expression that $T(\vec u)$ vanishes
in the semiclassical regime, as required by asymptotic freedom. This coincides
with \mw \lns\
in the appropriate cases.

Using the relation between
higher rank $SU(N)$ Yang-Mills theory and the Toda-Whitham hierarchy
\mor\marwar\naka\dw\don, one can introduce a set of  ``times" in the
prepotential which can be
seen
to be dual to the higher order Casimirs. This makes possible the computation of
contact terms for the two-observables coming from these Casimirs using the same
arguments
we have given here, and generalizing the expression \contactp\ to include the
rest of the time variables \gorsky. These variables were also considered in
\lns\ in the
context of the twisted theory, and the contact terms for the higher Casimirs
were derived using a blow-up argument.

\subsec{Remark on the normalization}

The overall normalization of the integral
\higheruplane\  has a meaning and can in principle
be fixed by physical computations or by
comparison to topological invariants.
In particular the constants $\alpha, \beta$ in
\afunction\bfunction\ are functions of the
group and for $G=SU(N)$ are functions of $N$.
Some constraints on these constants can be
obtained from the factorization of the measure
in certain regions of $\CM_{\rm Coulomb}$
expected on physical grounds.

Let us focus on $G=SU(N)$ and consider a
semiclassical region of moduli space with
scalar vevs:
\eqn\scalarvvp{
\phi = M \pmatrix{
N_2 {\bf 1_{N_1} } &  0 \cr
0 & - N_1 {\bf 1_{N_2} } \cr}
+
\pmatrix{ \phi_1 & 0 \cr 0 & \phi_2 \cr}
}
where $\phi_1, \phi_2$ are traceless. Since it is
important to keep track of quantum scales to understand
the behavior of the measure we introduce the
quantum scale $\Lambda_N$ and require that
$Z_{DW}$ be dimensionless.  We work in the
semiclassical region
\eqn\regionprm{
\vert M \vert
\gg \vert \phi_1^a - \phi_1^b \vert,
 \vert \phi_2^i - \phi_2^j \vert
\gg \vert \Lambda_N \vert
}
for $1\leq a < b \leq N_1$,
$N_1 +1 \leq i<j  \leq N_1+N_2$.
The physics of this region is
that we have a hierarchy of symmetry breakings:
\eqn\hierarchy{
SU(N) {\buildrel M  \over \longrightarrow  }
SU(N_1) \times SU(N_2) \times U(1)
\rightarrow U(1)^{N_1-1}
\times U(1)^{N_2-1} \times U(1)
}
with $N=N_1 + N_2$. At the large scale $M$
we integrate out $N_1 N_2$ vectormultiplets
corresponding to the off-diagonal blocks.
It is not difficult to
show that, up to relative corrections of order ${\cal O} (\phi/M)$,  the
semiclassical prepotential reduces to:
\eqn\semiclpp{
\eqalign{
\CF =  { i \over  4 \pi} \sum_{a<b}
(\phi_{1a} - \phi_{1b})^2
&  \log {(\phi_{1a} - \phi_{1b})^2 \over  \Lambda_{N_1}^2} \qquad\qquad\qquad
\cr
 +
{ i \over  4 \pi} \sum_{i<j  }
(\phi_{2i} - \phi_{2j })^2 \log {(\phi_{2i} - \phi_{2j })^2 \over
\Lambda_{N_2}^2} & + {i \over  4 \pi} N_1 N_2 (NM)^2 \log { (NM)^2 \over
\Lambda_N^2 }    \cr}
}
with renormalization group matching conditions:
\eqn\rgmatching{
{ \Lambda_{N_1} \over  \Lambda_N} =
\biggl( {\Lambda_N \over  NM}\biggr)^{N_2},
\qquad
{ \Lambda_{N_2} \over  \Lambda_N} =
\biggl({\Lambda_N \over  NM}\biggr)^{N_1}. \qquad
}
One can then show that  the $SU(N)$
$\Psi$-function with
scale $\Lambda_N$, denoted $\Psi_{SU(N), \Lambda_N} $
factorizes in the region \regionprm\ as:
\eqn\factorize{
\Psi_{SU(N), \Lambda_N} \rightarrow
\Psi_{SU(N_1), \Lambda_{N_1} }
\Psi_{SU(N_2), \Lambda_{N_2}} \Psi_{U(1)}  \Bigl( 1 + \CO(\phi/M)\Bigr)
}
Moreover, in the semiclassical region we have:
\eqn\factorab{
A^\chi B^\sigma \rightarrow \alpha(N)^\chi \beta(N)^\sigma
\biggl(\prod_{\vec \alpha  >0} \Bigl({ \vec \alpha\cdot \vec \phi \over
\Lambda_N} \Bigr) \biggr)^{(\chi + \sigma)/2}
}
with the product over the positive roots of
$SU(N)$. From this we easily find the factorization
in the region \regionprm:
\eqn\factorabp{
\prod_{\vec \alpha  >0} (\vec \alpha\cdot \vec \phi)
\rightarrow
(NM)^{N_1 N_2}
\prod_{\vec \alpha_1 >0} (\vec \alpha_1\cdot \vec \phi_1)
\prod_{\vec \alpha_2 >0} (\vec \alpha_2\cdot \vec \phi_2)
}
where $\vec \alpha_1, \vec \alpha_2$ are positive
roots of $SU(N_1), SU(N_2)$, respectively.

Thus, the nontrivial functions in the measure factorize in
the region \regionprm\ as expected on physical
grounds. Factorization of the entire measure implies that the measure
${d M  d {\overline M} \over  \vert \Lambda_{U(1)} \vert^2}$
picks up nontrivial dependence on $\vert M \vert^2$
which we have not predicted on {\it a priori} grounds.
However, the holomorphic part of the measure,
$(NM)^{N_1 N_2} $ can be expected
on {\it a priori} grounds
since it accounts   for the $R$-charge
anomaly of the vectormultiplets integrated out at
scale $M$. Combining this insight with Seiberg's trick of regarding constants
in an effective Lagrangian as vev's in some theory at
a higher scale to determine holomorphic dependence,
we can give an heuristic argument for
the $N$-dependence of $\alpha(N),\beta(N)$.
We regard the constants $\alpha(N)$, $\beta(N)$ as
well as $\Lambda_N$ as carrying $R$-charge.
Thus, as in \factorabp\
we expect the factorization
\eqn\alphafact{
\alpha(N) =
\alpha(N_1) \alpha(N_2) (\alpha_{U(1)})^{N_1 N_2}
}
for some constant $\alpha_{U(1)}$. Consequently, there should
be  $N$-independent constants $\kappa_1, \kappa_2 $ such that
\eqn\alphafact{
\alpha(N) = e^{\kappa_1 N + \kappa_2 N^2}
}

Similar formulae hold for $\beta$. Unfortunately we can
only fix one linear combination using the known
constants for   the $SU(2)$ case,
which have been found in \monopole\mw\ by
comparing to
explicit results for Donaldson invariants.
As remarked in
\wittk\monopole, to compare the results
of the physical theory to mathematical results, one has to
multiply the Donaldson-Witten function by the order
of the center of the gauge group.

\newsec{Single-valuedness of the integrand}

The generalized $u$-plane integral \higheruplane\ derived in the previous
section is
{\it not} manifestly well-defined because of monodromy around divisors
where the SW curve $\Sigma$ (or the abelian variety $J(\Sigma)$ in
the integrable system) degenerates. In this section we perform a careful check
of the monodromy invariance of the integral in the
case of simply-connected manifolds. The semiclassical analysis applies to any
simply-laced
gauge group. The strong-coupling analysis is only complete for $G=SU(3)$.

\subsec{Semiclassical monodromy}

The classical monodromy group is isomorphic to
the Weyl group of the gauge group, and it is
generated by the Weyl reflections $r_{i}$ associated
to the root basis,  $i=1, \cdots, r$. Semiclassically this monodromy has
a quantum correction due to the one-loop contribution
to the prepotential.

The general form of the semiclassical monodromy has
been presented in \klt \dansun\ for any gauge group. The action of the
$r_i$ monodromy on ${\vec a}$ is given
by the matrix
\eqn\weyl{
r_i = {\bf 1} - {\vec \alpha}_i \otimes
{\vec w}_i,
}
where the simple roots $\vec \alpha_i$ are expanded in
the Dynkin basis, and $\vec w_i$ are the fundamental weights.
The classical monodromy acting on
$({\vec a_D}, {\vec a})$ is given then by
\eqn\classmon{
P^{(r_i)}= \pmatrix{ (r_i^{-1})^t & 0 \cr 0 & r_i }.
}
The one-loop correction to the prepotential
\eqn\oneloop{
\CF_{\rm one-loop}={i \over 4\pi} \sum_{\vec \alpha>0} Z_{\vec \alpha}^2 \log
\bigl(
{ Z_{\vec \alpha}^2 \over \Lambda^2} \bigr),
}
(where the sum is over the positive roots) gives, in addition to
the Weyl reflection, a theta-shift in
the coupling constant of the form
\eqn\thetashift{
\tau \rightarrow (r_i^{-1})^{t} [\tau- {\vec \alpha}_i \otimes {\vec \alpha}_i]
r_i^{-1}.
}
The semiclassical monodromy matrix is then given by
\eqn\semimon{
M^{(r_i)}=\pmatrix{ (r_i^{-1})^t &  -(r_i^{-1})^t  ({\vec \alpha}_i \otimes
{\vec \alpha}_i)
\cr 0 & r_i }.
}
The invariance of the Coulomb path integral
 under these monodromies
is a non-trivial check of our expression.
First we analyze the lattice sum, then the measure factor
including the gravitational contributions.
To analyze the lattice sum, it is convenient to redefine the
variables $\lambda$, $D$, $\eta$ and $\chi$ by performing the
Weyl transformation $r_i^{-1}$. Notice that the lattice $\Gamma$ is invariant
under this redefinition, as the fundamental weights are shifted
by roots. The measure for the variables $D$, $\eta$, $\chi$ is invariant under
this transformation, since it is an orthogonal transformation. Also, the
two-observables $V_I$ are derivatives with
respect to $a_I$ of duality-invariant quantities, so they transform
as $\partial /\partial a_I$, therefore the terms involving the two-observables
remain invariant after the Weyl transformation of $\lambda$ and $D$.  For the
phase factor depending on $w_2(X)$, we can use the properties of the Cartan-
Killing form and the vector ${\vec \rho}$ and see that it gives the additional
term $\pi i (w_2(X), {\vec \alpha}_i \cdot  {\vec \lambda})$, where the dot
denotes the Cartan-Killing form on the weight space and $(\cdot, \cdot )$
denotes the usual product in integer cohomology.
We then have an additional phase factor in the lattice sum:
\eqn\add{
\exp \biggl( i \pi ({\vec \lambda} \cdot {\vec \alpha}_i, {\vec \lambda} \cdot
{\vec \alpha}_i)
+i\pi (w_2(X), {\vec \alpha}_i \cdot {\vec \lambda})\biggr).
}
The simple roots are expanded in the Dynkin basis. To see that this phase
factor is one, we take into account the decomposition in \eleclinebdle. The
term
\add\ then reads
\eqn\addexpl{
\eqalign{
& i \pi \biggl( ({\vec \lambda}_{\IZ} \cdot {\vec \alpha}_i, {\vec
\lambda}_{\IZ} \cdot
{\vec \alpha}_i)
+  (w_2(X), {\vec \lambda}_{\IZ} \cdot {\vec \alpha}_i ) +
\sum_{I,J=1}^{c-1} ({\vec m}_I \cdot {\vec \alpha}_i)({\vec m}_J\cdot {\vec
\alpha}_i )(\pi^I, \pi^J) \cr
& \,\,\,\,\,\ + \sum_{I=1}^{c-1}  ({\vec m}_I \cdot {\vec \alpha}_i) (w_2(X),
\pi^I)  + 2\sum_{I=1}^{c-1}  ({\vec m}_I \cdot {\vec \alpha}_i ) (\pi^I, {\vec
\lambda}_{\IZ} \cdot
{\vec \alpha}_i )\biggr). \cr}
}
The last term is an even integer, and  the other terms can be combined into
even integers
using the Wu formula
\eqn\Wu{
(w_2 (X), z) = (z,z) \,\ {\rm mod} \,\ 2, \,\,\,\ z \in H^2(X, Z),
}
Therefore, the lattice sum is invariant under the semiclassical monodromy.

Next we examine the measure factor in the Coulomb path
integral. The measure $[da d{\bar a}]$ is invariant under the monodromy,
and for the gravitational factor involving $\chi$ we can use
the symplectic transformation properties and the fact that
$\det r_i =-1$
to derive
\eqn\chitrans{
\biggl( \det {\p u_I \over  \p a^J} \biggr)^{\chi/2} \rightarrow \exp [{\pi i
\chi
\over 2}]\biggl( \det {\p u_I \over  \p a^J} \biggr)^{\chi/2}.
}
Finally, we analyze the factor involving the discriminant. In the semiclassical
regime we can use the expression \semidisc.
The Weyl reflection acts as follows on the $Z_{\alpha}$, with ${\alpha}>0$: the
basic
root $\alpha_i$ changes its sign, therefore
\eqn\ztrans{
Z_{\alpha_i} \rightarrow -Z_{\alpha_i},
}
and the rest of the positive roots are permuted, so the product
of the rest of the roots $Z_{\alpha}$ in the
classical discriminant is invariant. The only change in the
discriminant comes from this minus sign, and we finally obtain
\eqn\sigmatrans{
\Delta_{\Lambda}^{\sigma/8} \rightarrow \exp[ {i \pi \sigma \over 2}]
\Delta_{ \Lambda}^{\sigma/8}.
}
For a four-manifold with $b_1=0$ and $b_2^+=1$, $\chi + \sigma =4$, and
the measure factor does not change under the monodromy. Therefore,  the Coulomb
integral
is invariant under the semiclassical monodromies.

\subsec{Duality transformations}

To analyze the quantum monodromy, we have to consider the
duality transformations of the Coulomb integral in the appropriate
descriptions. To do this, we introduce the generalization of
the lattice theta function of \mw\ to the higher rank case
\eqn\hrtheta{
\eqalign{
&\Theta _{\Gamma}(\tau_{IJ}, \alpha_I, \beta^I; P, \xi_I)= \exp \biggl[ - i \pi
 (\alpha_I, \beta^I)
+ {\pi \over 2} \Bigl( \xi_{I,+} ({\rm Im} \tau)^{IJ} \xi _{J,+} - \xi_{I,-}
({\rm Im} \tau)^{IJ} \xi _{J,-} \Big)
\biggr] \cr
&\,\,\,\,\,\,\,\,\,\ \times \sum_{\lambda \in \Gamma} \biggl[ - i \pi
{\overline \tau} _{IJ} ({\hat
\lambda}^I_+, {\hat \lambda}^J_+)
-  i \pi \tau_{IJ} ({\hat \lambda}^I_-, {\hat \lambda}^J_-) - 2 \pi i  ({\hat
\lambda}^I,  \xi_I)
+ 2\pi i ({\hat \lambda}^I, \alpha_I) \biggr] ,\cr
}
}
where $ {\hat \lambda}^I = \lambda^I + \beta^I$. Notice that in the rank one
case we
recover the complex conjugate of the theta function introduced in \mw.

If we take
\eqn\xis{
\eqalign{
\xi_I &= {1 \over 2 \pi} V_I S_- + { {\sqrt 2} \over 16 \pi} {\overline{\cal
F}}_{IJK} \eta^J
\chi ^K   \omega, \cr
\beta^I &= \sum_{J=1}^{c-1}m_J^{~I} \pi^J , \,\,\,\,\,\,\ \alpha_I = {1 \over
2} w_2(X),
\,\,\,\ I = 1, \cdots ,r, \cr}
}
and consider $\lambda^I$ as the integer class $\lambda^I_{\IZ}$ introduced in
\eleclinebdle,
we see that the lattice sum \hirkthet\ can be written as
\eqn\psitheta{
\eqalign{
\Psi = & \exp \bigl[ {S^2 \over 8 \pi} V_I ({\rm Im} \tau)^{IJ}V_J \bigr]
\exp [ i \pi (\alpha_I, \beta^I)]  (\det {\rm Im}
 \tau_{IJ})^{-1/2}
\cr
& \times \int  \prod d \eta d \chi \exp \bigl[ { {\sqrt 2} \over 16 \pi}
{\overline {\cal F}}_{IJK}
\eta^I \chi ^J ({\rm Im} \tau)^{KL} V_L (S, \omega) \bigr]
\Theta_{\Gamma}  (\tau_{IJ}, \alpha_I, \beta^I; P, \xi_I).
\cr
}
}
The overall factor involving $S^2$ combines with $T(\vec u)$ to give the
duality-invariant
quantity ${\hat T}(\vec u)$ introduced in \dualct.

We will now consider the transformation properties of this
theta function under the group $Sp(2r, {\IZ})$. The generators
of the symplectic group are given in \spgens. The transformation properties are
the following:

Under $\Omega$ we have:
\eqn\omegatrans{
\eqalign{
&\Theta_{\Gamma} (- (\tau^{-1})^{IJ}, \alpha_I, \beta^I; P, -({\overline
\tau^{-1}})^{IJ} \xi_{J,+}
-( \tau^{-1})^{IJ} \xi_{J,-} ) \cr
&= {\sqrt { |\Gamma| \over |\Gamma'|}}
(\det i {\overline \tau}_{IJ})^{b_+/2} (\det  -i \tau_{IJ})^{b_-/2}
\Theta_{\Gamma'} (\tau_{IJ}, \beta^I, -\alpha_I; P, \xi_I).\cr
}
}
where $\Gamma'$ is the dual lattice. To derive the transformation law for
$\xi_I$ in \omegatrans, one has to use that
\eqn\imtaus{
({\rm Im} \tau)^{IJ} - 2i (\tau^{-1})^{IJ}= ({\rm Im} \tau)^{IK} {\overline
\tau}_ {KL} (\tau^{-1})^{LJ}.
}

If there is a characteristic element $w_2$ such that $(\lambda^I, w_2) =
(\lambda^I,
\lambda^I) \,\ {\rm mod} \,\ 2$, the transformation law of \hrtheta\ under
$T_{\theta}$ is:
\eqn\ttrans{
 \Theta_{\Gamma}  (\tau_{IJ}+ \theta_{IJ}, \alpha_I, \beta^I; P, \xi_I)=
\exp[ {\pi i \over 2}\sum_{I}  (w_2, \theta_{II} \beta^I)]  \Theta_{\Gamma}
(\tau_{IJ}, \alpha_I- {1\over 2}
\theta_{II}w_2 - \theta_{IJ} \beta^J,  \beta^I; P, \xi_I).
 }
Finally, under the transformation $\cal A$ we have
\eqn\atrans{
\Theta_{\Gamma} (A \tau A^{t}, \alpha_I, \beta^I; P, \xi^I)= \Theta_{\Gamma}
(\tau_{IJ}, A^{-1}\alpha, A^{t}\beta; P, A^{-1} \xi).
}

Using these transformations, it is easy to check that  the lattice sum
\psitheta (except for
the exponential involving $S^2$ and the phase) is then a modular form of
weights $((b_- +1)/2, (b_+-3)/2)$.
To derive this result, one formally considers the Grassmann variables $\eta$,
$\chi$ as
modular forms of weight $(0, 1)$, and takes into account the change induced in
the
Grassmann measure. The modular factors then combine with the measure $[da d
{\overline a}]$ and the gravitational factors to give the Coulomb integral for
the dual variables.

\subsec{Explicit check of quantum monodromy invariance for $SU(3)$}
Using the above transformation properties, we can analyze the quantum monodromy
in the $SU(3)$ case, as we know the explicit strong coupling spectrum in this
case \klt. To obtain the appropriate form of the integral, we will make a
symplectic
transformation for each pair of mutually local charges in such a way
that in the resulting theory there are two electrically charged particles
with charges $q_i^I = \delta_i^I$, $i, I=1,2$. For the two massless states
$\vec \nu_1$, $\vec \nu_2$ in \strongspec, we have to perform the
transformation $\Omega^{-1}$. For the states
$\vec \nu_3$, $\vec \nu_4$, the appropriate symplectic
transformation is given by $\Omega^{-1} {\cal A}^{-1} T_{-\theta}$, where
\eqn\sympone{
A= \pmatrix{ -1&-1 \cr
    1& 0\cr }, \,\,\,\,\,\ \theta=\pmatrix{ 1 & 1 \cr
                                                                    1&-2\cr}.
}
Finally, for $\vec \nu_5$, $\vec \nu_6$, the symplectic transformation has
again the structure $\Omega^{-1} {\cal A}^{-1} T_{-\theta}$ with
\eqn\sympone{
A= \pmatrix{ 0&1 \cr
    -1& -1\cr }, \,\,\,\,\,\ \theta=\pmatrix{ 2 & -1 \cr
                                                                    -1&-1\cr}.
}
Therefore, in this basis, the monodromies associated to the
two mutually local massless states are given by:
\eqn\strongmon{
M_i= \pmatrix{ {\bf 1} & {\bf e}^{ii} \cr
0 & {\bf 1} }, \,\,\,\,\ i=1,2,
}
where $({\bf e}^{ii})_{IJ}=\delta_I^i \delta_J^i$, and this holds for each of
the
three pairs of mutually local dyons. An important aspect of these
transformations is in that
in all cases we are left with dual theories where the shifts in the $\Gamma$
lattice are given by
\eqn\dualshfits{
\beta^I = {1 \over 2} w_2 (X), \,\,\,\,\,\ I =1, \cdots , r,
}
{\it i.e.} they are ${\rm Spin}^c$ structures. This result is obtained using
the higher rank
theta function transformations \omegatrans, \ttrans\ and \atrans. It is
important
to notice that the shifts in the $\Gamma$ lattice are defined
modulo integer cohomology classes.

Now the monodromy invariance of the integral
can be easily checked. The strong coupling monodromies are just
theta-angle
shifts in the dual coupling constants, and they are given by
\eqn\strongtheta{
\theta_{IJ}^{(i)}=\delta_I^i \delta_J^i.
}
We can now use \ttrans\ to see that the only change in the
higher rank theta function is given by the phase
$\exp [ -\pi i w_2(X)^2/4]$. There is also a change in the measure
associated to the factor involving the discriminant. Near a singular
locus this factor has the structure given in \strongdisc, and the monodromy
acts on $Z$ as $Z\rightarrow {\rm e}^{2\pi i} Z$. We then obtain a factor
$\exp[i \pi \sigma/4]$. But the second Stiefel-Whitney class verifies that
\eqn\sigmamod{
w_2 (X)^2 = \sigma \,\ {\rm mod} \,\ 8,
}
therefore both phases combine to $1$ and the integral is invariant under
the strong coupling monodromies.

\subsec{General case}
In the general case, the verification of quantum monodromy invariance requires
a precise knowledge of the strong coupling spectrum (or, equivalently, of the
monodromy subgroup of the symplectic group associated to the relevant
hyperelliptic
curve). On the other hand, one can always write the monodromy associated to
to a monopole divisor in the form \strongmon\ with a submatrix $q^2 {\bf
e}^{ii}$, where
$q= {\rm gcd} ( \nu_k)$, through an appropriate symplectic transformation \ad.
If
this symplectic
transformation is such that the $U(1)^i$ factor has a shift like \dualshfits,
then the above argument goes through. Notice that, near the divisor where a
hypermultiplet of charge $q$ becomes massless,  the discriminant of the curve
has a zero of order $q^2$. Thus
a general proof of quantum monodromy invariance follows if the symplectic
transformation taking the monodromy to \strongmon\  also
makes the dual line bundle $\lambda^i$ a ${\rm Spin}^c$ structure.
Unfortunately, the
monodromy group $\Gamma$ has not been studied in sufficient detail to make the
explicit check, although we fully expect it to work.

\newsec{Definition of the integral and wall-crossing}

The generalized $u$-plane integral \higheruplane\  above is a formal (albeit
monodromy-
invariant) expression. In order to give meaning to the integral and,
ultimately, derive
topological invariants for four-manifolds, we must define it carefully and
examine its metric
dependence properties.

\subsec{Defining the integral}

The integrand of \higheruplane\ has bad behaviour at the singularities on the
moduli
space. Therefore, we should regularize it appropriately near the codimension
one submanifolds
where dyons become massless and also near infinity.

The first step in the regularization is to choose appropriate coordinates along
these
submanifolds. A divisor where a hypermultiplet becomes massless can be given
locally by the
equation $a^i=0$, and this gives a preferred coordinate along this locus. The
other coordinates should be chosen according to the region we are
considering along the locus. At the ${\cal N}=1$ points, for instance, one
should
choose dyonic coordinates for all the dyonic $U(1)$ factors, while at the
region where a monopole locus goes to infinity, one should
choose electric coordinates for the remaining variables. At a generic point at
infinity, we choose electric
variables for all the $U(1)$ factors. Finally, near a point where mutually
non-local dyons become massless, there are no truly appropriate
coordinates, and the divergences of the quantities involved in the integral are
quite different from the ones associated to the monopole loci and to infinity.
In the case
of the AD points of the $SU(3)$ theory, we will check that the integral is
well-behaved near these points by using
the $\epsilon$, $\rho$ coordinates introduced in \adpar.

The second step in the regularization is to introduce appropriate cutoffs at
the singularities. In the case of the monopole loci, we choose tubular
neighbourhoods defined by $|a^i|>r$, where $r$ is
some radius that we will take to zero at the end. \foot{ We trust there will be
no confusion
with $r$ as the rank of the gauge group, nor $r_i$ as a Weyl reflection.} Near
a generic point at
infinity, the prepotential is naturally expressed in terms of the combinations
$Z_{\vec \alpha}$, and the
cutoff is given
by $|Z_{\vec \alpha}|<R$, where $R\rightarrow \infty$, for some positive roots
$\vec \alpha$ (different
choices of these roots give different directions at infinity). Near the AD
points of the $SU(3)$
theory, we will introduce an IR cutoff
in the $\epsilon$ plane, $|\epsilon|>r$, and analyze the behaviour of the
theory when we take
$r \rightarrow 0$ (recall that $\epsilon$ is a coordinate near this point).
Notice that this regularization can be interpreted as
the substitution of the original moduli space ${\cal M}_{\rm Coulomb}$ by a
``regularized"
moduli space ${\cal M}_{\rm Coulomb}^{\rm reg}$, which is a manifold with a
non-connected boundary.

Finally, we perform the integrals over the corresponding variables. The
procedure is now similar
to the one in \mw. We first perform the integral over the phase of the complex
coordinates chosen for each region, and this procedure gives a projection of
the terms of the form
$a^{\nu} {\overline a}^{\mu}$ onto terms with $\nu=\mu$. As we will see in the
next sections, in the
$SU(3)$ case, the resulting integrals converge, although their metric
dependence can
be discontinuous, resulting in wall-crossing.

\subsec{Metric dependence of the integral}

To study the possible metric dependence of the
integral, we follow the strategy in \mw. We consider
the variation of the Coulomb integral with respect  to
a first order
variation   $\delta\omega$ in the period point. All the
dependence on $\omega$ in the Coulomb integral
appears in the lattice sum $\Psi$. The variation is most easily expressed using
the
representation \findiml, and reads
\eqn\metvarone{
\eqalign{
\delta \Psi = &\sum_{\lambda \in \Gamma}
\int \prod_{I=1}^r d\eta^I d \chi^I \int_{-\infty}^{+\infty} \prod_{I=1}^r
d b^I  {\rm e}^{-S_E} \Bigl[  - 4 \pi ({\rm Im}  \tau_{IJ})\lambda_+^I
(\lambda^J , \delta \omega)
-{i \over 4 \pi} V_I (S, \delta\omega)b^I
\cr
&  -{i {\sqrt 2} \over 4} {\overline \CF}_{IJK} \eta^I \chi ^J (\lambda^K,
\delta\omega) +
 i V_I \bigl( (S, \delta\omega) \lambda_+^I + (\lambda^I, \delta\omega)S_+
\bigr)
\Bigr],\cr
}}
 where
\eqn\eucl{
\eqalign{
S_E =&
 i \pi {\overline \tau}_{IJ} (\lambda_+^I , \lambda_+^J)  +
i \pi \tau_{IJ} (\lambda_-^I , \lambda_-^J)
- {1 \over  8 \pi} b^I ({\rm Im } \tau)_{IJ} b^J + i V_I (S, \lambda_-^I) \cr
& +{i \over 4\pi} V_I
(S, \omega) b^I  +  { i \sqrt{2} \over  16 \pi} {\overline \CF}_{IJK} \eta^I
\chi^J
(b^K + 4 \pi
\lambda_+^K)  + i \pi ((\vec \lambda- \vec \lambda_0)  \cdot \vec \rho ,  w_2
(X)).\cr
}
}
Topological field theory promises us that $\delta \Psi$ is the integral of a
total derivative.
In fact, the metric variation in \metvarone\ can be written as
\eqn\metvartwo{
\delta \Psi = \sum_{\lambda \in \Gamma}
\int \prod_{I=1}^r d\eta^I d \chi^I \int_{-\infty}^{+\infty} \prod_{I=1}^r
d b^I  \Bigl[ \{ \overline \CQ, \Phi {\rm e} ^{-S_E} \} - 4 \pi {\partial \over
\partial b^J}
{\rm e} ^{-S_E} (\lambda^J, \delta \omega) \Bigr],
}
where
\eqn\var{
\Phi= i ({\rm Im} \tau)_{IJ} \chi^I  (\lambda^J, \delta \omega)  + {1 \over 4
\pi} (S, \delta \omega) V_I \chi^I.
}
Now we can use the fact that, according to the transformations \toptfin\
the $\overline \CQ$ operator is given by
\eqn\qbar{
\overline \CQ = i {\sqrt 2}  \eta^I {\partial \over \partial {\bar a}^I} + i (4
\pi
\lambda _{+}^I -b^I)
 {\partial \over \partial \chi^I}
}
and  write the metric variation as a total derivative in field
space with respect to the antiholomorphic coordinates:
\eqn\aholvar{
\delta \Psi = i{\sqrt 2}  {\partial \over \partial {\bar a}^I} \Upsilon^ {\bar
I} ,
}
where
\eqn\ups{
\Upsilon^{\bar I} = \sum _{\lambda \in \Gamma}
\int \prod_{J=1}^r d\eta^J d \chi^J \int_{-\infty}^{+\infty} \prod_{K=1}^r
d b^K \eta^I  \Phi{\rm e}^{-S_E}.
}
We introduce now an $(r, r-1)$ form on the Coulomb moduli space as
\eqn\form{
\Omega = i {\sqrt 2} \sum_{I=1}^r (-1)^{I+r-1} \Upsilon^{\bar I} da^1 \wedge
\cdots da^r \wedge
d{\overline a}^1 \wedge \cdots {\widehat { d {\overline a}^{\bar I} }  } \wedge
\cdots
\wedge d{\overline
a}^r.
}
which satisfies
\eqn\formeq{
d \Omega={\overline \partial} \Omega =
i {\sqrt 2} \bigl( \sum_{I=1}^r {\partial \over \partial {\overline a}^{\bar
I}} \Upsilon ^{\bar
I} \bigr) da^1 \wedge \cdots da^r \wedge d{\overline a}^1 \wedge \cdots \wedge
d{\overline a}^r. }
Taking into account that the measure and the observables in the Coulomb
integral
are holomorphic, we can use Stokes theorem to write the metric dependence
as an integral over the boundary of the regularized Coulomb branch:
\eqn\metric{
\delta Z_{\rm Coulomb}^{\rm reg} = \int_{\partial \CM_{\rm Coulomb}^{\rm reg} }
A^{\chi} B^{\sigma}
{\rm e}^{U + S^2 T_V} \Omega.
}

\subsec{Wall-crossing formulae along the monopole loci}

We will analyze now the generic wall-crossing at a monopole locus. For
simplicity and concreteness, we will focus on the case of the pure $SU(3)$
theory. The monopole locus is defined by the equation $a^2=0$. As explained
in section 4.3, there is a symplectic transformation to a basis in which
$\lambda^2 \in
H^2(X, {\IZ}) + {1\over 2} w_2(X)$. For the other $U(1)$ factor, according to
our remarks in
section 6.1, we should choose
the appropriate coordinates depending on the region of the monopole
locus, {\it i.e.} we are allowed to perform symplectic transformations that
leave $a^2$
fixed but change $a^1$.

Along the monopole locus, then, the behaviour of the $\tau_{22}$ coupling is
given by
\eqn\tautt{
\tau_{22} = {1 \over 2 \pi i} {\rm log} a^2 + \cdots, \,\,\,\,\,\ {\overline
\CF}_{222}=-{1 \over
2 \pi i} {1 \over {\overline a}^2} + \cdots,
}
while $\tau_{11}$, $\tau_{12}$ and the other ${\overline \CF}_{IJK}$ are smooth
 (except when we are at an ${\cal N}=1$ point or at
infinity, where we will have ``wall-crossing for wall-crossing". This will be
analyzed in a
moment). Denote $y \equiv {\rm Im}\tau_{22}$. An analysis similar to the one
performed in \mw\ shows that the possible discontinuities in the integral are
associated to terms involving only $1/(y^{1/2} {\overline a}^2)$, and they
occur when $(\lambda^2, \lambda^2)<0$, $\lambda_+^2=0$. These are the usual
conditions
for SW wall-crossing for $\lambda^2$.

Taking this into account, we can easily find the terms that contribute to
wall-crossing in the
integral \higheruplane, using the explicit expression in \rktwoint. First of
all, the factor
$\det {\rm Im} \tau_{IJ}$ appearing in the photon partition function has the
structure
\eqn\detim{
\det {\rm Im} \tau_{IJ}= y({\rm Im} \tau_{11} + O(1/y)),
}
and similarly
\eqn\taumatrix{
({\rm Im} \tau)^{-1}= \pmatrix{ ({\rm Im} \tau_{11})^{-1} + O(1/y) & O(1/y) \cr
                                                        O(1/y) & O(1/y)\cr}.
}
Therefore, in the term written in \rktwoint\ the only surviving contribution is
given by
\eqn\wcone{
-{1 \over 32 \pi } \lambda_+^2 {\overline \CF}_{111} {\overline \CF}_{222}
\bigl[ 4
\pi (\lambda^1,
\omega)
+ i ({\rm Im} \tau_{11})^{-1} V_1 (S, \omega) \bigr].
}
The term in $S_+^2$ (involving $({\rm Im} \tau)^{-1}$) in \hirkthet\ can
be analized in the same way, with the result that the only contribution comes
from $V_1^2 ({\rm Im} \tau_{11})^{-1} (S_+^2/8 \pi)$.  On the monopole locus,
we also
have the following expansion in powers of ${\overline a}^2$:
\eqn\taubarexp{
{\overline \tau}_{11}({\overline a}^1, {\overline a}^2) = {\overline
\tau}_{11}^{(0)} + {\overline a}^2
{\overline \tau}_{11}^{(1)} + \cdots,
}
where ${\overline \tau}_{11}^{(0)}={\overline \tau}_{11}({\overline a}^1,
{\overline a}^2=0)$ is the first term in an expansion in ${\overline a}^2$ and
is different from zero.
As we have already noticed, the positive powers of ${\overline a}^2$ do not
contribute
to the discontinuity of the integral, therefore we can put ${\overline
\tau}_{11}$ equal to
$ {\overline \tau}_{11}^{(0)} $ in the wall-crossing formula and write it as
the integral
of a residue:
\eqn\wcone{
\eqalign{
WC(\lambda^2) = & {{\sqrt 2} \over 8}{\rm e}^{i \phi (\lambda^2)}
\int_{{\cal D}} da^1 d{\overline a}^1 {\rm Res}_{a^2=0} \biggl\{
A^{\chi} B^{\sigma} \exp \bigl[pU + S^2 T_V-iV_2(S, \lambda^2) \bigr]
\cr
& \,\,\,\,\,\,\,\ \cdot \sum_{{\lambda}_1  \in \Gamma_1} q_{22}^{- (\lambda^2,
\lambda^2)/2 }
q_{12} ^{-( \lambda^1, \lambda^2)}
 \Psi (\lambda^1)\biggr\}. \cr}
}
In this expression, ${\cal D}$ denotes the monopole divisor,  $\phi(\lambda^2)$
is a global phase
which depends on $\lambda^2$ and is obtained through the appropriate
symplectic transformation to the monopole locus,   $q_{IJ}= \exp (2 \pi i
\tau_{IJ})$, and
\eqn\psione{
\eqalign{
 \Psi (\lambda^1) =  &{1  \over { \sqrt { ({\rm Im} \tau_{11})_{(0)} }} }
{\partial {\overline \tau}_{11}^{(0)} \over \partial {\overline a}^1}
\exp\bigl[   { 1 \over  8 \pi  }V_1^2 ({\rm Im} \tau_{11})^{-1}_{(0)} S_+^2
\bigr]
 \cr
&\cdot  \exp\biggl[ - i \pi {\overline\tau}_{11}^{(0)} (\lambda_+^1)^2
- i \pi   \tau_{11} (\lambda_-^1)^2
 - i   V_1  (S,\lambda_-^1)- i \pi (\lambda^1, \alpha_1) \biggr] \cr
& \cdot [4\pi (\lambda^1, \omega)
+ i ({\rm Im} \tau_{11})^{-1}_{(0)} V_1 (S, \omega) ] , \cr
}
}
where we have denoted $({\rm Im} \tau_{11})_{(0)}= (1/(2i))(\tau_{11} -
{\overline \tau}_{11}^{(0)})$ and $\alpha_1$ is the phase we have for the $a^1$
theory. This phase,
as well as the shift in the lattice $\Gamma_1$, depends on our choice of
symplectic basis
(we will make definite choices when we consider the ``wall-crossing for wall
crossing," because in
this case there is also a preferred $a^1$ variable).

Notice that the wall-crossing formula involves an integral which is very
similar to a rank one $u$-plane integral depending on a
``background field" $a^2$, and where the antiholomorphic part
of the theory involves a restriction to ${\overline a}^2=0$.

\subsec{Wall-crossing for wall-crossing}

An important aspect of the integral \wcone\ is that it has wall-crossing by
itself. Along the monopole locus, there are three distinguished points where
the $a^1$ theory
has singularities. These are the ${\cal N}=1$ points where the divisors
intersect, the
region at
infinity, and the AD points. The behaviour near the AD points will be analyzed
later.
In this section we will focus on the wall-crossing for wall-crossing near the
${\cal N}=1$
points and at infinity.

Near an ${\cal N}=1$ point the appropriate variable for the $a^1$ theory is
also a
``magnetic" one,
therefore near this point we have $a^1 \rightarrow 0$ and the behaviour of
$\tau_{11}$ is
similar to \tautt. For this choice of the variable, the shift in the lattice
$\Gamma_1$ is also
given by $w_2(X)/2$, {\it i.e.} the $\lambda^1$ are also ${\rm Spin}^c$
structures. The wall-crossing behaviour of the integral for the $a^1$ variable
is very similar
to the usual SW wall-crossing analyzed in \mw. Again,  we have wall-crossing
for $(\lambda^1, \lambda^1)<0$, $\lambda^1_+=0$. The discontinuity is now a
double residue and is given by
\eqn\wcwcsw{
\eqalign{
WC(WC(\lambda^1 , \lambda^2))= &- {4 \pi^2 }{\rm e}^{2\pi i (\lambda^I,
\lambda_0^I)}   {\rm Res}_{a^1, a^2=0} \biggl\{
A^{\chi} B^{\sigma} \exp \bigl[pU + S^2 T_V-iV_I(S, \lambda^I) \bigr] \cr
& \cdot \prod_{I,J=1}^2 q_{IJ}^{-{1\over 2} (\lambda^I, \lambda^J)}  \biggr\}
.\cr}
}
We have chosen the ${\cal N}=1$ point at $u= (27/4)^{1/3}\Lambda$. The phase
factor
involving
$\vec \lambda_0$ is the generalization to the higher rank case of a similar
factor considered in \mw. It gives the dependence of the SW contribution on the
generalized Stiefel-Whitney class, and can be obtained from \omegatrans. The
wall-crossing
for wall-crossing at the other ${\cal N}=1$ points can be obtained in a similar
way
(they will have
different global phases, according to the ${\IZ}_3$ symmetry).

At infinity along the monopole locus, the physics is that of an $SU(2)$ theory
embedded in $SU(3)$, {\it i.e.} we have the quantum-corrected gauge symmetry
breaking pattern $SU(3) \rightarrow U(1) \times SU(2)$, where the $U(1)$
(corresponding
to the $a^1$ theory) is weakly coupled in electric variables, and the $SU(2)
\rightarrow U(1)$
is weakly coupled in magnetic variables. There is a duality frame, therefore,
where
the behaviour of $\tau_{11}$ is given by
\eqn\tauoo{
\tau_{11} = {i \over 2\pi} {\rm log }a^1 + \cdots
}
and corresponds to electric variables, {\it i.e.} the shift in the lattice
$\Gamma_1$ is given
by $\beta^1 = (C^{-1}) ^1_{~J} \pi^J$.  The wall-crossing of the integral on
$a^1$ will then be
a Donaldson wall-crossing, exactly like the one anlayzed in \mw. The expression
we get
is formally identical to the one in \wcwcsw, although the conditions for
wall-crossing
in $\lambda^1$ are the ones for Donaldson wall-crossing, and one must use the
appropriate
duality frame.

\subsec{Wall-crossing at infinity}

The relevant information to analyze the wall-crossing at infinity is encoded in
the semiclassical one-loop
correction to the prepotential \oneloop. In the $SU(3)$ case it is given by:
\eqn\oneloopthree{
\CF_{\rm one-loop}={i \over 4\pi} \sum_{i=1}^3 Z_i^2 \log \bigl(
{ Z_i^2 \over \Lambda^2} \bigr),
}
where we denote $Z_i=Z_{{\vec \alpha}_i}$, corresponding to the three positive
roots of $SU(3)$, ${\vec \alpha}_i$, $i=1,2,3$. The explicit expressions are
$Z_1=2a^1-a^2$, $Z_2=-a^1+2a^2$, $Z_3=Z_1+ Z_2=a^1+a^2$.To analyze the
conditions  for wall-crossing, we focus on the photon partition function
of the lattice sum \hirkthet:
\eqn\expinf{
\exp\bigl[ - i \pi {\overline \tau}_{IJ} (\lambda_+^I , \lambda_+^J)  -
i \pi \tau_{IJ} (\lambda_-^I , \lambda_-^J)\bigr] \sim \prod_{\vec \alpha>0}
\biggl({{\overline Z}_{\vec \alpha} \over \Lambda}\biggr)^{-(\vec \lambda_+
\cdot \vec \alpha)^2}
\biggl({Z_{\vec \alpha} \over \Lambda}\biggr)^{(\vec \lambda_- \cdot \vec
\alpha)^2}.
}
We can approach the region at infinity in moduli space in many ways, keeping
one of the
$Z_i$, $i=1,2,3$ to be finite and the other two, $Z_j$, $j\not=i$, going to
infinity (notice that we cannot
keep two of the $Z_i$ finite, as $Z_3=Z_1+Z_2$). The conditions for a possible
wall-crossing in
$\vec \lambda$ are then given by $\vec \lambda_+\cdot \vec \alpha_j=0$, $j
\not=i$,
as one
can easily check from \expinf. As any two
positive roots are linearly independent, we find $\vec \lambda_+=0$. Therefore,
there is no
wall-crossing at infinity for $SU(3)$ (across codimension one walls): the
integral is not discontinuous when $\vec \lambda_+=0$.
This is in contrast with the case of the non-simple rank two group $SU(2)\times
SU(2)$,
where
there are only two positive roots and therefore there are directions at
infinity where one
finds wall-crossing (namely, the Donaldson wall-crossing associated to each of
the $SU(2)$ factors).

One can also check this behaviour for $SU(3)$ using the $u$, $v$ variables,
going to infinity
along the $u$ or the $v$ planes, and using the explicit expressions for the
behaviour of
the prepotential given in \klt. Again, one finds that  the condition for a
possible wall-crossing along these directions is $\vec \lambda_+=0$ and there
is no discontinuity in the integral.

\newsec{The blowup formula}

The blowup formula generalizing
\finstern\ can be easily derived following
the method used in  \mw. Since there are
manifolds with vanishing SW contributions
it suffices to derive the formula for
$Z_{\rm Coulomb}$. The latter is
easily derived by studying the change of
the measure in \higheruplane.  One then
applies a universality argument.

Let $\tilde X = Bl_P(X)$
be the blowup at a smooth point. Then
$\tilde \sigma = \sigma-1, \tilde \chi = \chi+1$.
The change in the measure under
$X \rightarrow \tilde X$ is just:
\eqn\chgmeas{
\mu_{\tilde X} = {\alpha \over  \beta}
\biggl( \det {  \p u_J \over  \p a^I} \biggr)^{1/2} \Delta_\Lambda^{-1/8} \mu_X
}
Now let  $B$ denote the class of the exceptional
divisor, with $B^2=-1$.  In the chamber
$B_+=0$ (or more properly, for a fixed correlation function,
where $B_+ < \epsilon$ for some sufficiently small
$\epsilon$) the $\Psi$ function factorizes to
a $\Psi$-function for $X$ times a holomorphic
$\Psi$-function involving a sum over the root lattice.
Indeed we may write:
\eqn\latvcts{
\eqalign{
\tilde \lambda^I_+ & = \lambda^I_+ \cr
\tilde \lambda^I_- & = \lambda^I_- + n^I B\cr}
}
where $n^I$ is in $\IZ+m_J^{~I} e^J $.
for integer $e^J$. The shift  $e^J$ depends
on the generalized Steifel-Whitney class of
 the gauge bundle   $\tilde E \rightarrow \tilde X$.
In the chamber $B_+=0$  the $\Psi$-function factorizes as:
\eqn\psifun{
\Psi_{\tilde X} = \sum_{n^I} e^{i \pi \tau_{IJ} n^I n^J
   + i t V_I n^I - i \pi \sum_I n^I} \Psi_X
\equiv
 \Theta_{ m^t \vec e  , \vec \Delta}(t \vec V \vert \tau) \Psi_X
}
where we have written  $\tilde S = S + t B$
and $\vec \Delta = (1,\dots, 1)$.
Thus, accounting for the contact term, the integrand
for the blown-up manifold $\tilde X$ is related to that
for $X$ by the replacement of zero-observables:
\eqn\hrblwi{
e^{U} \rightarrow e^U
{\alpha \over  \beta}
\biggl( \det {  \p u_J \over  \p a^I} \biggr)^{1/2} \Delta_\Lambda^{-1/8}
e^{-t^2 T_V}
 \Theta_{ m^t \vec e, \vec \Delta}(t \vec V \vert \tau)
}
Note that the expression must be monodromy invariant.
Indeed, it has modular weight zero.
This observation can be used to derive the required
contact terms $T_V$ for $V$ other than the
quadratic Casimir \lns.
Moreover, since it is invariant, it is a function of
$t$ and the Casimirs $u_2, \dots, u_r$.

Physically, we expect the defect $B$ creating
the blown-up manifold can be represented by
an infinite number of local observables. The ring of local BRST invariant
observables
is generated by the Casimirs $u_2, \dots, u_{r+1}$. Thus
there must be {\it polynomials} $\CB_{\vec e,k}(u_2, \dots, u_{r+1})$
such that
\eqn\hrblwii{
{\alpha \over  \beta}
\biggl( \det {  \p u_J \over  \p a^I} \biggr)^{1/2} \Delta_\Lambda^{-1/8}
e^{-t^2 T_V}
 \Theta_{ m^t  \vec e, \vec \Delta}(t \vec V \vert \tau)
= \sum_{k\geq 0} t^k \CB_{\vec e,k}(u_2, \dots, u_{r+1})
}
The fact that $\CB_{\vec e,k}(u_2, \dots, u_{r+1})$ are polynomials
can be proven as follows: the blowup expression \hrblwi\ is monodromy
invariant, in particular of weight zero, so it must be a
function of $u_I$, $I=2, \dots, r+1$, and
$t$. Using the $R$-symmetry, we see that $t$ has to be of charge $-2$, hence
the polynomial $\CB_{\vec e,k}$ has charge $2k$. On the other hand, the
expression
\hrblwii\ has no singularities in the moduli space. This is because the
theta function involved in the expression never has singularities, and the only
possible singularities come from $\Delta_{\Lambda}^{-1/8}$. But these must be
cancelled by zeros of the theta function, as follows from monodromy invariance.

In the case of the $SU(2)$ theory, the explicit expression for these
polynomials was obtained
in \mw\ using the expansion of the theta functions in terms of Eisenstein
series, but in the
higher rank case these expansions are not available. However, these
expressions can probably be obtained using the relation between
Seiberg-Witten theory and integrable systems. For $A_r$, the
integrable system relevant to the Seiberg-Witten solution is the periodic Toda
lattice
\mor \marwar \naka. The solutions to both
models are straightline motions in the Jacobian of a
hyperelliptic curve. Indeed, we recognize that \hrblwii\ is
essentially the $\tau$
function for the Toda hierarchy. Solutions to the
Toda equations can be obtained from the Baker-Akhiezer function,
and comparing the $t$ expansion of these solutions
should determine the polynomials
$\CB_{\vec e,k}(u_2, \dots, u_{r+1})$. We have not carried
out the details of this procedure.

In any case, the blowup formula at higher rank is:
\eqn\blwvi{
\eqalign{
\biggl\langle \exp\bigl[ I(S) + t I(B) + p \CO \bigr] \biggr\rangle_{\tilde{X}
}
& =   \biggl\langle
\exp\bigl[ I(S) +
 p \CO \bigr]   \tau(t \vert \CO_2, \dots , \CO_{r+1}) \biggr\rangle_{
X  } \cr
  = \sum_{k\geq 0}  t^k
\biggl\langle
\exp\bigl[ I(S) +
&
 p \CO \bigr]   \CB_{\vec e, k} (\CO_2, \dots, \CO_{r+1}) \biggr\rangle_{ X  }
\cr}
}

\newsec{Behaviour at the Argyres-Douglas points}

The Coulomb integral \higheruplane\ depends on the metric of the four-manifold
$X$. Its
variation with respect to the metric can be written in terms of an integral
over the
boundary of the regularized Coulomb branch, as in \metric. In general, this
integral over the boundary will vanish, due to the damping factors
associated to the behaviour of the couplings near the singularities or in the
semiclassical region. However, at the AD points of the $SU(3)$ theory, there is
an
${\cal N}=2$ superconformal field theory with a finite value of the gauge
coupling.
The situation
is reminiscent of the behaviour of the $N_f=4$ theory analyzed in \mw, where
it was found that generic correlation functions have a continuous dependence on
the
metric. Therefore, one should analyze the possible
{\it continuous} metric dependence associated to these
superconformal points.

\subsec{A general argument}

The blowup formula
derived in  section 7   severely constrains the
possibility of continuous metric dependence.
 This is because the blow-up
formula relates
the Donaldson-Witten function of manifolds with different signatures. As we
will show below, for sufficiently large
signature (e.g., $\sigma>-11$ for $G=SU(3)$) the
measure near the superconformal points is sufficiently
smooth that the metric variation vanishes. Now, the
blow-up formula relates the invariants on $\tilde X$ to
invariants on a manifold with
$\sigma(X) = \sigma(\tilde X) + 1$.  If there is no
continuous variation in the latter correlators there
cannot be any such variation in the former.
Care should be taken with this argument since the
blowup formula only applies for $\omega$
in certain chambers of the forward light cone
in $H^2(X;\IR)$.
For any given correlator, the formula applies in a
chamber with $B_+< \epsilon$ for some sufficiently
small $\epsilon$, where $B$ is the exceptional
divisor of the blow-up.
If there is no continuous metric variation in this chamber
then, given metric-independent
wall-crossing formulae, there cannot be any
continuous variation in any other chamber.
(In fact, as we have seen there is no wall-crossing from infinity
on codimension one walls, so there is really
 only one chamber.)

One could ask why an argument like this doesn't rule out
continuous metric dependence in the $N_f=4$ theory
considered in \mw. The reason is that, in this case, the inequality
involving the signature also includes the ghost number $Q$ of the
correlators, and the condition not to have any metric
dependence has the form of an {\it upper} bound on $2 \sigma + Q$.
The above argument does not apply in this case, and one can
easily check that the blowup formula is perfectly compatible
with continuous metric dependence for the $N_f=4$ theory. The
reason for the different behaviours (and for the different bounds on the
signature) has to do with the fact that, in the $N_f=4$ theory, the
continuous metric dependence comes from the behaviour
at infinity, while the superconformal points in $SU(N)$ super
Yang-Mills theories are in an ``interior" region of the moduli space.

\subsec{An explicit check}

The above argument is rather general and should be
checked by explicit computation. We now give a
detailed analysis of the behavior near the AD points
for $G=SU(3)$. In particular we explicitly show the
absence of continuous variation for $\sigma > -11$.

\subsubsec{Convergence near the AD points}

First of all,   we must analyze the convergence of the Coulomb
integral itself, as the divergences of the integrand of \higheruplane\ near the
superconformal point are rather different from the ones we have considered. We
have to introduce a cutoff $r$ for the variable $\epsilon$ introduced in
\adpar,
and study the behaviour of the integral as $r$ goes to zero, as we have
indicated
in section
6.1.  To do that, we first consider the antiholomorphic terms with
${\overline \epsilon}^{-n}$ behaviour. These come from the terms in ${\overline
\CF}_{IJK}$,
in \rktwoint, and whose structure near the AD point was presented in \prepotad.
The most
divergent term corresponds to ${\overline \CF}_{111}{\overline \CF}_{221} \sim
{\overline
\epsilon}^{-4}$. We have to write the measure of the integral in terms of
$\epsilon$, $\rho$ variables. The jacobian of the change of variables from
$a^I$ to $x^J=
\epsilon$, $\rho$ can be computed at leading order from \asad, \atad:
\eqn\jacoad{
  \det {\partial a^I \over \partial x^J}= { 6c \epsilon^{7/2} \over
\Lambda^{5/2}} H(\rho) + \cdots
}
The measure is then
\eqn\measure{
[da d{\overline a}]= \biggl( {36 c^2 \over \Lambda^5} |H(\rho)|^2 |\epsilon|^7
+ \cdots \biggr)
d\epsilon d{\overline \epsilon} d\rho d {\overline \rho}.
}
Because of the factor $|\epsilon|^7$ in the measure, we see that the
leading behaviour of the integral
is smooth, so it converges. Notice that that the rest of the terms involved
in the integrand
($V_I$, $\tau_{IJ}$,  $T_V$, $u$, $v$) are
smooth as $\epsilon$ goes to zero,
as one can see from \adpar, \adperiods, and \adtaus.
Thus we conclude that the integral is well-defined in
the limit $r \rightarrow 0$.

\subsubsec{Explicit formulae for the metric variation}

Now we want to study the possible metric dependence of the integral. The first
step in doing that is to write explicit expressions for the
$\Upsilon^{\overline I}$ quantities
defined in \ups. After doing the Grassmann integrals in the rank two case, one
obtains
\eqn\upstwo{
\eqalign{
\Upsilon^{\overline 1} =& -{i {\sqrt 2} \over {16 \pi} } (\det {\rm Im}
\tau)^{-1/2}
 \exp\bigl[   { 1 \over  8 \pi  }V_J ({\rm Im} \tau)^{JK} V_K  S_+^2 \bigr] \cr
& \cdot  \sum_{\lambda \in \Gamma }
 \exp\biggl[ - i \pi {\overline\tau}_{IJ} (\lambda_+^I,\lambda_+^J)
- i \pi   \tau_{IJ} (\lambda_-^I,\lambda_-^J)
- i \pi ((\vec \lambda  -{\vec \lambda}_0)  \cdot  \vec \rho,  w_2 (X))  - i
V_I  (S,\lambda_-^I)
\biggr] \cr
&\cdot [4\pi (\lambda_{+}^K, \omega)
+ i ({\rm Im} \tau)^ {KL} V_L (S, \omega) ]  \cr
&\cdot \biggl\{   {\overline \CF}_{22K} \bigl( i ({\rm Im}\tau)_{1J}
(\lambda^J, \dot \omega)
+ {1 \over 4 \pi} (S, \dot \omega)V_1 \bigr) -{\overline \CF}_{21K} \bigl( i
({\rm Im}\tau)_{2J} (\lambda^J, \dot \omega)
+ {1 \over 4 \pi} (S, \dot \omega)V_2 \bigr)  \biggr\} ,\cr
\Upsilon^{\overline 2} =& -{i {\sqrt 2} \over {16 \pi} } (\det {\rm Im}
\tau)^{-1/2}
 \exp\bigl[   { 1 \over  8 \pi  }V_J ({\rm Im} \tau)^{JK} V_K  S_+^2 \bigr] \cr
& \cdot \sum_{\lambda \in \Gamma }
 \exp\biggl[ - i \pi {\overline\tau}_{IJ} (\lambda_+^I,\lambda_+^J)
- i \pi   \tau_{IJ} (\lambda_-^I,\lambda_-^J)
- i \pi ((\vec \lambda  - {\vec \lambda}_0) \cdot  \vec \rho,  w_2 (X))  - i
V_I  (S,\lambda_-^I)
\biggr] \cr
&\cdot [4\pi (\lambda_{+}^K, \omega)
+ i ({\rm Im} \tau)^ {KL} V_L (S, \omega) ]  \cr
&\cdot \biggl\{ {\overline \CF}_{11K} \bigl( i ({\rm Im}\tau)_{2J} (\lambda^J,
\dot \omega)
+ {1 \over 4 \pi} (S, \dot \omega)V_2 \bigr) -{\overline \CF}_{12K} \bigl( i
({\rm Im}\tau)_{1J} (\lambda^J, \dot \omega)
+ {1 \over 4 \pi} (S, \dot \omega)V_1 \bigr)  \biggr\} .\cr}
}
To analyze the behaviour near the superconformal point, we use the duality
frame specified by
the symplectic transformation \symplead, in order to use the ``small torus"
\rhocurve\ and
the explicit solutions in section 2. The differential form $\Omega$ of \form\
can be
written now in terms of the $\epsilon$, $\rho$ variables.
The explicit expression follows from:
\eqn\omegaad{
\eqalign{
\Omega= &i {\sqrt 2} \bigl( \det {\partial a^I \over \partial x^J} \bigr)
\biggl\{  \bigl( { \partial {\overline a}^2 \over \partial {\overline \epsilon}
}\Upsilon^{\overline 1} -
{ \partial {\overline a}^1 \over \partial {\overline \epsilon}}
\Upsilon^{\overline 2}\bigr)
d\epsilon \wedge d\rho \wedge d {\overline \epsilon} \cr
& +  \bigl( { \partial {\overline a}^2 \over \partial {\overline \rho}}
\Upsilon^{\overline 1} -
{ \partial {\overline a}^1 \over \partial {\overline \rho}} \Upsilon^{\overline
2}\bigr)
d\epsilon \wedge d\rho \wedge d {\overline \rho} \biggr\}.\cr
}
}
There are two terms in \omegaad\ which can lead
to variation $\delta Z_{\rm Coulomb} $.
In the first term in \omegaad\ we
take the integral over the $\rho$ boundary,
which will be a set of three
tubular neighbourhoods of the monopole divisors $\rho^3=1$. The contributions
of these
boundaries leads to {\it discontinuous}, wall-crossing type,  metric
dependence. This    is
just the monopole wall-crossing analyzed in section 6.3.

The second term in \omegaad\ is more
interesting and it gives the possible metric
dependence associated to the AD points.
We regularize the integral by cutting a small disk
of radius
$r$ around $\epsilon =0$. The boundary integral
in $\epsilon$ will   then  be along the
circle of radius $r$, $S_r$, with center at $\epsilon=0$.
We want to know if
there are surviving contributions as $r \rightarrow 0$.

To analyze the integral over $S_r$ it is important to take into account
monodromy invariance
under $\epsilon \rightarrow {\rm e}^{2 \pi i } \epsilon$.
This   invariance can
be verified explicitly
using the fact, crucial to the entire argument we
are giving, that after the symplectic
transformation \symplead, the line bundles
$\lambda^1$ define ${\rm Spin}^c$ structures.
First, one can easily
check, using the behaviour of ${\overline \CF}_{IJK}$ in \prepotad, that all
the powers of ${\overline \epsilon}$
appearing in the expression
are positive or zero. Actually the only contribution
one can have when $r \rightarrow 0$ comes from the terms with no powers of
${\overline \epsilon}$. These involve ${\overline \CF}_{111}$, ${\overline
\CF}_{112}$.
We can write the metric dependence then as the integral of a residue, in the
same
way that we have written the wall-crossing formulae:
\eqn\metricdep{
\eqalign{
& \delta Z_{\rm Coulomb}(\omega )= -{1 \over 8 \pi}  \oint d \epsilon \int
d\rho d{\overline
\rho} A^{\chi} B^{\sigma}
 \biggl( \det {\partial a^I \over \partial x^J} \biggr) {\rm e}^{U+S^2 T_V}\cr
& \cdot  (\det {\rm Im} \tau _{(0)})^{-1/2}
 \exp\bigl[   { 1 \over  8 \pi  }V_J ({\rm Im} \tau)^{JK} _{(0)} V_K  S_+^2
\bigr]
\sum_{\lambda \in \Gamma }  \bigl[ i ({\rm Im}\tau)^{(0)}_{2J} (\lambda^J,
\delta \omega)
+ {1 \over 4 \pi} (S, \delta \omega)V_2 \bigr]  \cr
&  \cdot \exp\biggl[ - i \pi {\overline\tau }(\overline \rho)  (\lambda_+^1)^2-
i \pi
{\overline \tau}_{22}^{(0)}(\lambda_+^2)^2
- i \pi   \tau_{IJ} (\lambda_-^I,\lambda_-^J) - i   V_I  (S,\lambda_-^I)
\biggr] \cr
& \cdot  {d \overline \tau (\overline \rho) \over d {\overline \rho}} [4\pi
(\lambda_{+}^1, \omega)
+ i ({\rm Im} \tau)^ {1L} _{(0)} V_L (S, \omega) ], \cr
}
}
where the $(0)$ (sub)superscript means that in the antiholomorphic quantities
we take ${\overline \epsilon}=0$.
In \metricdep\ we have omitted a global phase
depending on the non-abelian magnetic fluxes.

The expression \metricdep\ is not zero in general.
We conclude that $Z_{\rm Coulomb}$ has continuous
metric dependence from the AD points.

\subsubsec{The $\rho$-plane theory}

We now examine the metric dependence we
have discovered in more detail.
One of the interesting things about
\metricdep\  is that it involves, essentially,
a rank one integral associated to the
elliptic curve \rhocurve. We will
refer to this curve as the ``$\rho$-curve.''
 To see this, let us study  the leading behaviour for
$\epsilon \rightarrow 0 $ of the measure appearing in \metricdep. Up to a
constant that can be
computed from  \suthreedisc,
\detb, and \jacoad, together with \period, we find the behaviour
\eqn\admeasure{
A^{\chi} B^{\sigma}
 \biggl( \det {\partial a^I \over \partial x^J} \biggr)  \sim \epsilon^{{3
\sigma + \chi + 14 \over 4}}
\Delta_{\rho}^{\sigma/8} \omega_{\rho}^{1- {\chi\over 2}},
}
where $\omega_\rho$ is the period of the curve
\rhocurve.
Similarly, using \adperiods\ we also have for the 2-observable
\eqn\twobsad{
V_1 \sim {\epsilon^{1/2} \over \omega_{\rho}},
}
which again behaves as the 2-observable of the rank one case (involving the
period of the $\rho$-curve). Comparing
the factors \admeasure\twobsad\ to the general expressions
for the rank one $u$-plane integrals we see that the
leading behavior for $\epsilon \rightarrow 0$ is governed by
a family of effective supersymmetric
theories described by the $\rho$-curve and which
we will refer to as the  ``$\rho$-plane theory.''
\foot{
One must  excercise caution when expressing the
behavior of the integral at the AD points in terms of the
$\rho$-plane theory since
the matrix $({\rm Im} \tau)^{IJ}$ and $(\det {\rm Im}
\tau)^{1/2}$ does lead to
 subleading terms in $1/({\rm Im} \tau (\rho))$.}

\subsubsec{Monopoles to the rescue}

The nonvanishing
continuous metric variation \metricdep\
of $Z_{\rm Coulomb}$
appears to spell doom for the topological
invariance of $Z_{DW}$. Before jumping to this
conclusion we must consider the possible
continuous metric variation of the other
terms in \fullform.  In particular, we must
examine the continuous dependence of the
mixed SW/Coulomb integrals along the monopole divisors
$Z_{\CD_i^{(1)}} $. In the present case the
relevant divisors are   ${\CD}_i^{(1)}$, $i=1,2,3$  defined
by the roots of  $\rho^3=1$ .
The integrals  $Z_{\CD_i^{(1)}} $ will be analyzed in some
detail in the next section, but their continuous metric dependence is easy to
analyze here. The Seiberg-Witten
contributions are obtained by cancellation of wall-crossing of $Z_{\rm
Coulomb}$ along the monopole divisors, and they are integrals along these
subvarieties involving the Seiberg-Witten invariants
at the singularities of the $\rho$-plane (corresponding to the dyons becoming
massless at
$\rho^3=1$). They are   obtained from the behaviour of $Z_{\rm Coulomb}$ near
$\rho^3=1$
in such a way that wall-crossings cancel:
\eqn\cancelwall{
WC_{\rho_i} (Z_{\rm Coulomb})+ WC( Z_{\CD_i^{(1)}} ) = 0
}
where $\rho_i$, $i=1,2,3$
are the roots of $\rho^3=1$ and label the three monopole divisors near the AD
point.
We want to know the
continuous metric dependence of these Seiberg-Witten
contributions. That is we want to compute
${ \delta \over  \delta \omega} Z_{\CD_i^{(1)}} $
for generic $\omega$, not just at walls.
The continuous variation comes from the region
$\epsilon=0$, and we will denote this variation by
  $\delta_{\epsilon=0}Z_{\CD_i^{(1)}} $ .
Since the continuous metric
dependence and the discontinuous metric dependence involve
the behaviour with respect to different variables, we see that the
wall-crossing of the integral
over $\rho$, $\overline \rho$ in \metricdep\ near the $\rho^3=1$ divisors has
to match the Seiberg-Witten wall-crossing of
$\delta_{\epsilon=0}Z_{\CD_i^{(1)}} $ at these
singularities. We then have
\eqn\metdepsw{
\delta_{\epsilon=0}Z_{\CD_i^{(1)}} = \oint d\epsilon \epsilon^{ {3 \sigma +
\chi + 14} \over 4}
\sum_{\lambda \in \Gamma} SW (\lambda^1) {\rm Res}_{\rho=\rho_i} F(\rho,
\epsilon, \lambda^I, \delta \omega) ,
}
 where $F(\rho, \epsilon, \lambda^I, \delta \omega)$ is a holomorphic function
of $\epsilon$, $\rho$ which depends also on $\lambda^I$ and $\delta\omega$. It
can be obtained, as we have
indicated, by computing the wall-crossing of the $\rho$, $\overline \rho$
integral in \metricdep\ and matching it to the wall-crossing of a
Seiberg-Witten contribution with
the appropriate insertion of observables,
as in the following section.
\foot{ Notice once more that the consistency of this procedure requires that
the $\lambda^1$ bundles, which are the ``line bundles'' that couple to the
$\rho$ theory,
define ${\rm Spin}^c$ structures.}  Now we note  that in the $\rho$-plane
integral in \metricdep, all the terms that do not correspond to a rank one
integral for the curve \rhocurve\ do not contribute
to wall-crossing, as they involve subleading powers in $1/({\rm Im}
\tau(\rho))$. Thus, the continuous metric dependence
of $Z_{\CD_i^{(1)}} $ is
expressed in terms of the $\rho$-plane theory.

Taking this into account, the metric
dependence of $Z_{DW}$ near the AD point
is a sum of two terms: one
from the   integral in \metricdep\ and
one from the Seiberg-Witten contributions
near the singularities at $\epsilon=0, \rho^3=1$, and can
therefore be written schematically as
\eqn\metricdeptot{
\delta Z_{DW} =  \oint d \epsilon
\epsilon^{ {3 \sigma + \chi + 14} \over 4} \Bigl\{
\int d\rho d{\overline \rho}~ [\cdots] +
\sum_{\lambda \in \Gamma} \sum_{i=1}^3 SW(\lambda^1) {\rm Res}_{\rho=\rho_i}
F(\rho, \epsilon, \lambda^I, \delta \omega) \Bigr\},
}
where $[\cdots]$ denotes the integrand of \metricdep\ up to the global power
of $\epsilon$ that we have factored out.

\subsubsec{Vanishing of $\delta Z_{DW}$ for
$G=SU(3)$, $\sigma>-11$.}

We are finally ready to justify the assertion that
$\delta Z_{DW}=0$ for sufficiently large signature.
This is a simple consequence of
\metricdeptot. From the  scaling behaviour of the terms in
\metricdep\ we see that all of the
terms in the $\epsilon$ expansion of \metricdep\ have positive powers.
Therefore,  \metricdeptot\ will vanish if the power of $\epsilon$ in the
measure is bigger than $-1$, {\it i.e.}, if
\eqn\nomet{
\sigma > -11,
}
 where we have taken into account that $\chi + \sigma=4$. Notice that we can
always make
insertions of 2-observables which have no leading powers of $\epsilon$ (for
example,
$V_2 = \Lambda /c + O(\epsilon)$). Therefore, we can not write a general
selection rule
involving the ghost number of a given correlator, as in the $N_f=4$ case
analyzed in
\mw. Rather we have a condition on the signature of the manifold, given by the
bound
\nomet. This bound is particular to the gauge group $SU(3)$. For other
superconformal points associated to other gauge groups and/or
matter content \apsw \eguchisc, we expect
other explicit bounds depending on the $R$-charge spectrum near these points.

We can now complete the  argument for topological
invariance of $Z_{DW}$ by invoking the general
argument at the beginning of this section since
the blow-up formula holds for $Z_{DW}$ and the measure factor depending on
$\epsilon$ is common to both contributions
in \metricdeptot.

\newsec{The Seiberg-Witten contributions}

\subsec{The SW contribution along the monopole loci}

As in \mw, we expect that the higher rank Donaldson-Witten functional is
given by the Coulomb integral \higheruplane\ plus the contributions coming from
the
monopole divisors, as we have indicated in \fullform. Generically, along theses
divisors a dyon becomes massless
and the low-energy effective theory contains one hypermultiplet coupled to one
of
the $U(1)$ factors.  The twisted theory is now a ``mixed" theory where one of
the variables
(the one that we have called $a^2$) is a distinguished coordinate but we can
still perform
duality transformations which leave $a^2$ fixed. We expect, however, that
the twisted
theory will localize to supersymmetric configurations for the $A^2$ vector
multiplet
coupled to the hypermultiplet. These simply give the Seiberg-Witten monopole
equations
for the $A^2$ variables. At the ${\cal N}=1$ points, there are distinguished
coordinates
for both $a^1$, $a^2$, the effective theory contains two mutually local
hypermutiplets (each of
them coupled to each of the vector multiplets), and the twisted theory will
localize to supersymmetric configurations for {\it both} vector multiplets
coupled to the hypermultiplets.
At these points, the contribution will be given then by Seiberg-Witten
invariants $SW(\lambda^1)$, $SW (\lambda^2)$.

On general grounds, the Donaldson-Witten functional for $SU(3)$ will be given
by
\eqn\zdw{
\eqalign{
Z_{DW} = &Z_{\rm Coulomb} + \sum_{i} \int_{{\cal D}^{(1)}_i} da_i^1
d{\overline a}_i^1 \sum_{\lambda^1, \lambda^2} \int_{{\cal M}_{SW}(\lambda^2) }
\mu^i_{\lambda^1, \lambda^2} (a^1_i, {\overline a}^1_i, a^2) \cr
& + \sum_{i=1}^3  \sum_{\lambda^1, \lambda^2} \int_{{\cal M}_{SW}(\lambda^1)
\times {\cal M}_{SW} (\lambda^2)}
\Phi_{i} (a^1, a^2),\cr}
}
where we have included a sum over the
components of the
codimension one divisor ${\cal D}^{(1)}_i$, and also the
contribution
of the three ${\cal N}=1$ points. ${\cal M}_{SW}(\lambda)$ is the
Seiberg-Witten
moduli space for the ${\rm Spin}^c$ structure $\lambda$.The structure of the
functions  $\mu^i_{\lambda^1,
\lambda^2} (a^1_i, {\overline a}^1_i, a^2)$, $\Phi_{i} (a^1, a^2)$ can be
obtained by cancellation of
wall-crossing, as in \mw, and comparing to the
formulae derived in section 6.  We find that the Seiberg-Witten contribution
along a
monopole divisor is given by the function
\eqn\mufunc{
\eqalign{
 \mu^i_{\lambda^1, \lambda^2} (a^1_i, {\overline a}^1_i, a^2)=&{\rm e}^{\phi_i
(\lambda^2)} \exp \bigl[pU +
S^2 T_V-iV_2(S, \lambda^2) \bigr]  C_{22}(a^1, a^2)^{(\lambda^2)^2/2}
C_{12}(a^1, a^2) ^{(\lambda^1, \lambda^2)} \cr
& \cdot P(a^1, a^2) ^{\sigma/8} L(a^1, a^2) ^{\chi/4} \Psi (\lambda^1), \cr}
}
where $\Psi(\lambda^1)$ is given in \psione, $\phi_i (\lambda^2)$ is the
appropriate
global phase depending on the divisor and the corresponding symplectic
transformation, and the functions $C_{22}$,
$C_{12}$, $P(a^1, a^2)$, $L(a^1, a^2)$ are given by
\eqn\functions{
\eqalign{
C_{22}(a^1, a^2)& = { a^2 \over q_{22}}, \cr
C_{12}(a^1, a^2) & = q_{12}^{-1}, \cr
L(a^1, a^2)&=-{{\sqrt 2} \over 8} \alpha^2 \bigl( \det {\partial u_I \over
\partial a^J}
\bigr)^{2},\cr
P(a^1, a^2)&= {1 \over 32} \beta^{8} { \Delta_{\Lambda} \over a^2} .\cr
}
}
As we explained in section 8, the Seiberg-Witten contributions along the
monopole divisors
have continuous metric dependence near the AD point, which can be obtained by
matching
the wall-crossing of the $\rho$, $\overline \rho$ integral in \metricdep\ near
$\rho^3=1$ to the wall-crossing coming from the Seiberg-Witten
contributions at these singularities. This can be verified
using the    computations above, with
the only difference that instead of having an integral over $a^1_i$, ${\bar
a}^1_i$  (the coordinate which parametrizes the monopole divisors) we have a
contour integral in $\epsilon$.

\subsec{Contributions from the ${\cal N}=1$ points}

Now we follow the same approach to compute the functions involved at the ${\cal
N}=1$
points.
By comparison with wall-crossing, their structure is
\eqn\phinone{
\eqalign{
\Phi_i (a^1, a^2)= &e^{i \phi_i} \exp [ 2\pi i (\lambda^I, \lambda_0^I) ] \exp
\bigl[pU + S^2 T_V-iV_I(S, \lambda^I) \bigr]  \cr
& \cdot \prod_{I,J=1}^2\bigl( {\widetilde C}_{IJ} (a^1, a^2) \bigr)
^{{ 1 \over 2} (\lambda^I, \lambda^J)}  {\widetilde P} (a^1, a^2) ^{\sigma/8}
{\widetilde L} (a^1, a^2) ^{\chi/4},\cr}
}
where $\phi_i$ is a global phase depending on the generalized Stiefel-Whitney
class and on the ${\CN}=1$ point. The functions ${\widetilde C}_{IJ}(a^1,
a^2)$, $I$, $J=1,2$, ${\widetilde
P}(a^1, a^2)$, ${\widetilde L}(a^1, a^2)$ are given by
\eqn\functionsnone{
\eqalign{
{\widetilde C}_{II}(a^1, a^2)&={a^I \over q_{II}}, \,\ I =1,2 \cr
{\widetilde C}_{IJ}(a^1, a^2)& = {q_{IJ}^{-1}},  \,\ I,J=1,2, \,\  I \not=J,
\cr
{\widetilde L}(a^1, a^2)&= -4 \pi^2 \alpha^2 \bigl( \det {\partial u_I \over
\partial
a^J} \bigr)^{2},\cr
{\widetilde P}(a^1, a^2)&= 16 \pi ^4 \beta^{8} { \Delta_{\Lambda} \over a^1
a^2} .\cr
}
}

We can thus write the SW contribution at the ${\cal N}=1$ points for $SU(N)$,
which
is
a straightforward generalization of the above procedure,
\eqn\swnone{
\eqalign{
\langle {\rm e} ^{U + I_2(S)} \rangle_{\lambda^1, \cdots, \lambda^r}^{(i)} =&
\alpha^\chi \beta^\sigma {\rm e}^{i \phi_i} {\rm e}^{ 2\pi i (\lambda^I,
\lambda_0^I)}
\Bigl(\prod_{I=1}^r
SW(\lambda^I)\Bigr)  \cr
& \cdot {\rm Res}_{a^1= \cdots a^r=0}
\biggl\{ \Bigl( \prod_{I=1}^r (a^I)^ { {2 \chi + 3 \sigma \over 8}-{
(\lambda^I)^2\over 2}-1}
 {\widetilde q}_{II}^{-(\lambda^I)^2/2}  \Bigr) \prod_{1\le I< J \le r}
\Bigl( q_{IJ}^{-(\lambda^I, \lambda^J)}  \Bigr)  \cr
&\cdot \biggl( { \Delta_{\Lambda}
\over \prod_{I=1}^r a^I} \biggr)^{\sigma/8} \bigl(
\det {\partial u_I \over \partial a^J} \bigr)^{\chi/2}  \exp \bigl[ U + S^2
T_V-iV_I(S, \lambda^I) \bigr]\biggr\} ,\cr
}
}
where ${\widetilde q}_{II}= q_{II}/a^I$, and we have included in $\alpha$,
$\beta$ the numerical factors that are obtained, as in \functionsnone, from
matching to wall-crossing.
It is important to notice that the
quantities ${\widetilde q}_{II}$ as well as the factor involving
$\Delta_{\Lambda} /\prod_{I=1}^r a^I$ in \swnone\ are regular at $a^I=0$.

\subsec{$SU(N)$ Donaldson invariants for manifolds of simple type}

In this section we generalize the gauge group
to $G=SU(N)$ for
all $N$, but specialize the class of manifolds to
those of simple type.

In the simple type case, we can evaluate the contribution at the ${\cal N}=1$
points
using the explicit
expressions given in \ds \dpstrong\ for the ${\CN}=1$ point where $N-1$
monopoles become massless, together
with the discrete ${\IZ}_{4N}$ symmetry relating the ${\CN}=1$ vacua. The
eigenvalues for $\phi$ are \ds\
\eqn\eigen{
\phi_n = 2 \cos {\pi (n-{1 \over 2}) \over N}, \,\,\ n=1, \cdots, N.
}
and from this expression we can easily compute the VEVs of the Casimirs,
\eqn\vevs{
c_{2s} \equiv \langle {\rm Tr} \phi^{2s} \rangle = {2s \choose s} N, \,\,\,\,\
c_{2s+1} \equiv  \langle {\rm Tr} \phi^{2s+1} \rangle=0,
}
One also finds a relation
\eqn\relation{
\sum_{I=1}^r {\partial a^I \over \partial \phi_i} \sin {\pi k I \over N} =
i \cos {\pi k (i-{1 \over 2}) \over N},
}
and from this we can obtain,
\eqn\twonone{
{\partial u_2 \over \partial a^I}= -2i \sin {\pi I \over N}.
}
This gives the value of $V_I$.

Finally, we need the value of the off-diagonal couplings at
the ${\cal N}=1$ point. These can be obtained using the scaling trajectory for
the eigenvalues
$\phi_n$ obtained in \ds. This trajectory depends on a parameter $s$, and $s=0$
corresponds
to the ${\cal N}=1$ point. The value of the magnetic gauge coupling along this
trajectory is given
by:
\eqn\scaling{
\tau_{IJ}(s)={2 \over N}  \sum_{K=1}^{N-1} \tau_{K}(s) \sin {\pi I K \over N}
\sin {\pi J K \over N},}
and the eigenvalues of this matrix, $\tau_K(s)$, can be explicitly written in
terms of
some integrals. In particular, the leading behaviour of $\tau_K (s)$ as $s
\rightarrow 0$ is given by \ds
\eqn\integral{
\tau_{K}(s)={ i \over 2 \pi \sin {\pi \kappa \over 2 } } \int_{-b}^b  d\theta
{\cos (1-\kappa)\theta
\over {\sqrt {{\rm e}^{-2s} - \sin^2 \theta} }},}
where $\kappa=K/N$ and $b=\arcsin {\rm e}^{-s}$. This integral has a divergent
part ${1 \over 2\pi i } \log s$ as $s\rightarrow 0$, and this produces the
usual logarithmic divergence, diagonal
in $I$ and $J$, in $\tau_{IJ}$. But the off-diagonal terms of $\tau_{IJ}$ are
finite at $s=0$ ({\it i.e.} at the ${\cal N}=1$ point), and they can be
obtained from the finite part of the integral in \integral\ For $SU(3)$ we have
for instance \klt,
\eqn\offsuthree{
\tau_{12}(0)={i \over \pi} \log 2,}
as one can check from \integral\ and \scaling. Although we have not
found an explicit expression for the finite part of \integral,
there are two important properties of the
couplings $\tau_{IJ}(0)$, $I \not=J$,  that one can deduce from  \integral\ and
\scaling: they are imaginary, and they satisfy the following symmetry property
\eqn\symoff{
\tau_{IJ}(0)=\tau_{N-I \,\ N-J}(0).
}

We can already write the
contribution from the
${\cal N}=1$ points to the
$SU(N)$ invariants, using the fact that these points are related by the
${\IZ}_{4N} \subset
U(1)_R$ symmetry. We must take into account the $R$-charges of the different
operators
in the correlation function, as well as the gravitational contribution to the
anomaly that appears
on a curved four-manifold. This anomaly can be computed from the microscopic
theory
(as in \wittk) or directly from the expression given in \swnone. The two
computations
must give the same result because the factors in the measure were actually
determined
from an $R$-charge argument. In fact, as the $R$-charge of $q_{II}$ is zero, we
have
\eqn\macrorcharges{
\eqalign{
R \biggl( { \Delta_{\Lambda} \over \prod_{I=1}^r a^I} \biggr)^{\sigma/8}=
&{\sigma \over 2}  N(N-1)-{\sigma \over 4} (N-1),\cr
R  \bigl(
\det {\partial u_I \over \partial a^J} \bigr)^{\chi \over 2} = & {\chi \over 2}
N(N-1),\cr
R \Bigl( \prod_{I=1}^r{\widetilde q}_{II}^{-(\lambda^I)^2/2}  \Bigr)=& (N-1){
2\chi + 3 \sigma \over 4},\cr
}
}
and the $R$-charges of these terms give the right $R$-charge coming from the
underlying
twisted SYM theory, namely
\eqn\anomaly{
{N^2-1 \over 2} (\chi + \sigma).
}
We conclude
that the contribution of the $\CN=1$ vacua is
\eqn\simple{
\eqalign{
&\langle {\rm e} ^{U + I_2(S)} \rangle_{SU(N)} = {\widetilde \alpha}_N^\chi
{\widetilde \beta}_N^\sigma \sum_{k=0}^{N-1} \omega^{k[(N^2-1)
\delta + N \vec \lambda_0\cdot \vec \lambda_0]} \sum_{\lambda^I} {\rm e}^{ 2\pi
i
(\lambda^I, \lambda_0^I) }  \Bigl(\prod_{I=1}^{N-1}
SW(\lambda^I)\Bigr)\cr
& \,\,\,\,\,\  \cdot  \prod_{1\le I< J \le r}
\Bigl( q_{IJ}^{-(\lambda^I, \lambda^J)}  \Bigr) \exp \biggl[
\sum_{s=1}^{[{N-1\over 2}]}
p_{2s}\omega^{2ks} c_{2s} + 2\omega^{2k} S^2 + 2 \omega^k \sum_{I=1}^{N-1} (S,
\lambda^I)  \sin {\pi I \over N} \biggr],\cr}
}
where $\omega=\exp[i \pi /N]$ and $\delta=(\chi + \sigma)/ 4$.  In $\widetilde
\alpha$,
$\widetilde \beta$ we have reabsorbed numerical factors that come from the
evaluation of $L$, $P$ and $\widetilde q_{II}$ at the point where $N-1$
monopoles become massless (the
values at the other points
are obtained using $R$-symmetry.) The $q_{IJ}$, $I<J$, can be obtained from
\scaling\ and \integral\ when $s=0$, as we have discussed. We have also
included in the
phase factor labeling the ${\CN}=1$ vacua in \simple\ an additional term
depending on
the
generalized Stiefel-Whitney class, which generalizes the $SO(3)$ case
considered
in
\wittk \monopole. This term can be obtained if we take into account that the
instanton
number of the bundle, once non-abelian fluxes are included, satisfies
\instanton. Equivalently, one can take into account the transformations
\omegatrans, \ttrans and \atrans\ to find this extra factor
when we go through the different ${\CN}=1$ vacua and
see that they generate precisely this extra phase.

We will now find a natural generalization of the phase factor ${\rm e}^{i \pi
w_2^2(E)/2}$ obtained in \mw\ to guarantee that the resulting expression is
real. Notice that this is also a
consistency check of the above answer, as this factor must be a global phase
depending
on the generalized Stiefel-Whitney class. We then have to consider the
properties of \simple\ under complex conjugation (assuming that the overall
factor
$ {\widetilde \alpha}_N^\chi {\widetilde \beta}_N^\sigma$ is real). First
notice that the first term
in the sum, $k=0$, changes by conjugation of the global phase:
$\exp [-2\pi i (\lambda^I, \lambda_0^I)]$ (the factors $q_{IJ}$, $I<J$, are
real). Now we take into account that, due to
\symoff\ and the expression \twonone, the right hand side of \simple\ has the
symmetry $\lambda^I \rightarrow \lambda^{N-I}$, hence we can write the
resulting phase as
\eqn\realphases{
{\rm e}^{ 2\pi i  (\lambda^I, \lambda_0^I)}
{\rm e}^{-2\pi i [(\lambda^I, \lambda_0^I) + (\lambda^{N-I}, \lambda_0^I)]}.
}
Using the fact that $(C^{-1})_J^{~I} + (C^{-1})_J^{~N-I}$ is
an integer, for all $I,J=1, \dots, N-1$,  we can write the
second  factor as
\eqn\last{
\exp \biggl\{ -\pi i\sum_{J=1}^{N-1} [J(J+N-2)(\pi^J, w_2(X))]
\biggr\}=\exp[\pi i  N \vec v \cdot \vec v ] ,
}
where we have taken into account the Wu formula \Wu\ and the explicit form of
the
inverse Cartan matrix for $SU(N)$. For the rest of the terms in the sum, $k=1,
\dots, N-1$, we take into account that, under conjugation,
$\omega^k \rightarrow -\omega^{N-k}$, and we change $\lambda^I \rightarrow
-\lambda^I$. Using the transformation \monopole:
\eqn\swflip{
SW(-\lambda) = (-1)^{\delta}SW(\lambda),
}
one easily checks that the sum of the terms $k=1, \dots, N-1$ changes by an
overall sign of the form
$(-1)^{N \vec v \cdot \vec v}$ (notice that, for manifolds of simple type,
$\delta$ is an integer). Comparing with \last\ we see that, under conjugation,
\simple\
picks a global sign depending on the generalized Stiefel-Whitney class $\vec
v$. Notice that
$N \vec v \cdot \vec v$ is always an integer. Moreover, for $N$ odd it is an
even integer, because in this case $NC^{-1}$ is an even form. Therefore, for
$N$
odd, \simple\ is real. For $N$ even, it is then natural to
include a phase factor of the form ${\rm e}^{i \pi N \vec v \cdot \vec v/2}$ to
make the above
expression real. This factor is independent of the lifting of $\vec v$ as long
as $N$ is even, and
in the special case of  $SU(2)$ we recover the factor introduced in \mw.

\newsec{Application 1: Twisted ${\cal N}=2$ superconformal field theories}
At the AD points there is an ${\CN}=2$  superconformal field theory, and as
we are studying the twisted version of ${\CN}=2$ super Yang-Mills theory, the
relevant spacetime symmetry algebra describing the model there is the twisted
version
of the ${\CN}=2$ extended superconformal algebra in four dimensions.
Recall that this algebra includes extra bosonic generators $K_{\mu}$ and $D$,
corresponding to the special conformal transformations and the dilatations,
respectively,
as well as two Weyl spinors $S_{A I}$, ${\overline S}^{\dot A I}$, where $A$,
$\dot A$
are spinor indices of $SU(2)_{-} \times SU(2)_+$, and $I$ is the $SU(2)_R$
index. There is also an $R$ generator corresponding
to the non-anomalous $U(1)_R$-current. The topological twist changes the
coupling to gravity of
all the fields charged with respect to the internal symmetry group $SU(2)_R$,
in the usual
way. This means that in the twisted theory we consider the diagonal subgroup
$SU(2)'$ of $SU(2)_+ \times SU(2)_R$, and the internal symmetry index $I$ is
promoted to a spinor index, $I \rightarrow \dot A$. We then obtain two scalar
supercharges,
\eqn\sccharges{
{\overline {\CQ}}= \epsilon^{\dot A \dot B} {\overline Q}_{\dot A \dot B},
\,\,\,\,\,\
{\overline S}= \epsilon^{\dot A \dot B} {\overline S}_{\dot A \dot B},
}
where ${\overline Q}_{\dot A I}$ is the usual supersymmetry charge. We can also
define descent operators,
\eqn\scdescent{
G_{\mu}= {i \over 4} {\overline \sigma}_{\mu}^{\dot B  A} Q_{A \dot B},
\,\,\,\,\,\
T_{\mu}= {i \over 4} {\overline \sigma}_{\mu}^{\dot B  A} S_{A \dot B}.
}
The twisted ${\CN}=2$ superconformal algebra includes the relations:
\eqn\twsca{
\eqalign{
[\overline{\CQ}, D ]  =  {1\over 2} {\overline {\CQ}} ,
\quad & \quad
[\overline{S}, D]  =  -{1 \over 2} \overline S,\cr
[\overline{\CQ}, R]  = -{\overline {\CQ}}, \quad & \quad
[\overline{S}, R]  =  \overline S ,\cr
[G_{\mu}, D]  = {1 \over 2} G_{\mu}  , \quad & \quad
[T_{\mu}, D]  =-{1 \over 2}T_{\mu},  \cr
[G_{\mu}, R]  = G_{\mu},   \quad & \quad
[T_{\mu}, R]= -T_{\mu},\cr
\{ {\overline {\CQ}}, G_{\mu} \}= i P_\mu, \quad & \quad
\{ {\overline S}, T_{\mu} \}=iK_{\mu},\cr
\{ {\overline {\CQ}} , {\overline{\CQ}} \}=0, \quad & \quad
\{ \overline {S}, {\overline{S}} \}=0, \cr
\{ {\overline {\CQ}}, {\overline S} \}=&2R-4D.
\cr}
}
One can define in a natural way (topological) chiral primary fields as those
fields satisfying
\eqn\chipri{
\{ {\overline {\CQ}}, \Phi ] =\{ {\overline S}, \Phi]=0.
}
{}From the last relation in \twsca, we recall the well-known fact that for such
fields $R(\Phi)=2D(\Phi)$. The
topological descendants of a chiral primary field are $n$-forms with the
structure,
\eqn\desc{
\Phi^{(n)}_{\mu_1 \cdots \mu_n}=\{ G_{\mu_1}, \cdots, \{ G_{\mu_n}, \Phi]
\cdots ],
}
and we see from \twsca\ that they are also annihilated by ${\overline S}$.
After integrating them
on $n$-cycles, we find new chiral primary fields. Notice that the $R$-charge of
the
topological descendant $\Phi^{(n)}$ satisfies $R(\Phi^{(n)})=R(\Phi)-n$.

Now we can try to extract some information about the twisted superconformal
field
theory at the AD point from the results we have obtained above. An important
new
feature arising near the AD points is that the gravitational
factors $A^{\chi}$, $B^{\sigma}$ have the following behaviour:
\eqn\gravad{
A^{\chi} B^{\sigma} \sim \epsilon^{\chi/4} \epsilon^{3\sigma/4}.
}
Recall that this factor measures the gravitational contribution to the anomaly
of the $R$-current.
The factor involving $\sigma$ is naturally interpreted as the $U(1)_R$ anomaly
of the
three mutually nonlocal hypermultiplets becoming massless at the AD point. We
interpret the
factor involving $\chi$ as a signal of the $R$ charge of the
superconformal vacuum, leading to an anomaly
$- \chi(X)/4$ in units where the anomaly of a
single hypermultiplet is $- \sigma(X)/4$.
We conclude that in the
twisted superconformal theory on a manifold $X$
there is a selection rule for correlation functions:
\eqn\selection{
\langle \Phi_1 \cdots \Phi_n \rangle_X \not= 0
}
only for
\eqn\selectp{
\sum_i R[\Phi_i] = {1\over 10}\chi(X),
}
where we have taken into account that $R(\epsilon)=2/5$.
This has a striking ressemblance to the
selection rule for correlators in a twisted
$d=2$ $\CN=(2,2)$ sigma model
on a Riemann surface $\Sigma$, where the $R$-charge of the vacuum is given by
$-{\hat c}\chi(\Sigma)/2$.

The generalization of \selection\selectp\ to $SU(N)$
can be determined by examining the order of vanishing
of $\det {\p a \over  \p u} $.  For simplicity, assume $N$ is odd. Since $u_j
\sim \epsilon^j$,
and there are ${N-1 \over 2}$ vanishing $\beta$-periods, $a_i \sim
\epsilon^{N+2 \over 2}$,
we expect that
$\det {\p a \over  \p u} \sim \epsilon^{-(N-1)^2/8} $ and
hence the RHS of \selectp\ becomes
 $ {1 \over 8}{(N-1)^2 \over N+2} \chi(X)$ for
the $\IZ_N$ multicritical superconformal theories.
\foot{In deriving this result we have simply
counted factors of $\epsilon$ in the determinant
$\det {\p a \over  \p u} $. In principle the $\CO(1)$ factors
in this determinant, which we have not computed,
could lead to a cancellation of the coefficient
of the leading
divergence  $\epsilon^{-(N-1)^2/8}$. We are assuming
this does not happen.   }
It is interesting to compare this result with some
similar recent results for $\CN=1$ theories
\freedman.

We believe that further information about the behavior of superconformal
theories can be extracted from the above results, in
particular from the $\rho$-plane theory of section 8
\mmprep.

\newsec{Application 2: Large $N$ limits of $SU(N)$
$\CN=2$ SYM}

As an application of \simple\ we now sketch the
large $N$ asymptotics of the Donaldson invariants
for $SU(N)$. While this has no obvious interest for
topology, explicit results for correlators in supersymmetric
Yang-Mills theory are rarities. It is even rarer that
one can explicitly study a  large $N$ limit using exact
results.   Since on hyperk\"ahler manifolds the correlators of $\overline
\CQ$-invariant operators are the same  in the  topologically twisted theory as
the ``physical correlators''   \vw\ we may hope to understand
something of the physics of large $N$ $\CN=2$
$SU(N)$ SYM theory. We know from \ds\ that
this limit presents some unusual features.

Te folllowing identities will be useful to evaluate the correlators from
\simple:
\eqn\simpsum{
\eqalign{
\sum_{k=0}^{N-1} \omega^{k \ell} &
= N \qquad\qquad \ell = 0 ~ \mod ~ 2N,\cr
& = 0 \qquad\qquad \ell = 0 ~  \mod ~ 2 , \qquad
\ell\not=0 ~ \mod ~ 2 N \cr
& = {2 \over  1- \omega^\ell} \qquad\qquad \ell = 1 ~ \mod ~ 2.\cr}
}
The last case, $\ell$ odd,
 leads to a nontrivial $1/N$ series:
\eqn\oddexp{
\sum_{k=1}^{N-1} \omega^{k \ell}
= {2 i \over  \pi \ell} N + 1 + 2 i \sum_{t=1}^\infty {B_{2t} (-1)^t \over
(2t)!} \bigl({\pi \ell \over  N}\bigr)^{2t-1}
}
where the $B_{2t}$ are Bernoulli numbers.

\subsec{The torus $X=T^4$. }

Although our considerations in this paper have been mostly on
simply-connected manifolds, the extension to the non-simply connected case
can be done along the same lines (some interesting issues
arising in this case are adressed in
\nsc). For the four-torus, however, the situation is
very simple because the only basic class is $\lambda=0$, and
the Donaldson invariants are still given by \simple\ (with $\lambda=0$).
The reason for this is the following: since $T^4$ is flat,
the monopole field in the SW
equations must vanish and the SW moduli space
is simply the space of harmonic 1-forms on $T^4$.
However, this is a nongeneric situation and the
complications of the bundle of antighost zeromodes
are most easily handled, as is standard, by perturbing
the equations. Then, since
 $T^4$ is hyperkahler the only basic class is
$\lambda=0$, with $SW(\lambda) = 1$.  This is
in accord with the physical argument using a
nowhere-vanishing mass perturbation \wittk.

Let us consider the operators in the theory.
Since the torus is not simply connected we could
also introduce
 $\overline \CQ$- closed 1-cycle and 3-cycle operators.
These only contribute through their contact terms
and do not change the following results in any
essential way, so we omit these operators.
It is convenient to rescale the $0$-observables to
\eqn\rescale{
A_j \equiv {1 \over  c_{2j} }  {\Tr} \phi^{2j}
}
for $j=1,2,\dots$. Then we have, simply,
\eqn\simpletor{
\Biggl\langle \exp\biggl( \sum t_j A_j + I(S) \biggr) \Biggr\rangle
= \sum_{k=0}^{N-1} \exp\bigl[\sum_{j \ge 1} t_j \omega^{2kj} + 2 S^2
\omega^{2k} \bigr].
}

Now we must decide how to define the large $N$ limit.
We consider the $N \rightarrow \infty$ limit
of  finite polynomials in $t_j$ and $S$. By \simpsum\ we
see that {\it all correlators at fixed ghost number vanish
identically for sufficiently large $N$.} The large $N$
limit exists, but it is utterly trivial.

We can obtain some more interesting correlators in
two ways. The first is to add $SU(N)/\IZ_N$ fluxes to the
theory. These produce a factor $\omega^{-k f}$ in the
sum over $\CN=1$ vacua, where
$f = - N \vec v^2$, $\vec v= \sum_{I=1}^{N-1} \pi^I \vec w_I$ with
$\vec w_I$ the fundamental weights and
$\pi^I$ integral classes. A second way to get interesting
correlators is to introduce a new ``conjugate'' set of
operators:
\eqn\conjops{
\bar A_j \equiv { 1 \over  c_{2L+2-2j} } {\Tr} \phi^{2L + 2 - 2j},
}
where $L \equiv [(N-1)/2] $ and $j=1,2,\dots$.
Note that these operators do not have well-defined
ghost number in the $N \rightarrow \infty$ limit.

With these modifications \simpletor\ becomes a little
more intricate:
\eqn\simpletori{
\eqalign{
\Biggl\langle \exp\biggl( \sum t_j A_j + \sum \bar t_j \bar A_j+ I(S) \biggr)
\Biggr\rangle_f
& = \qquad\qquad  \qquad\qquad\cr
 \sum_{k=0}^{N-1} \omega^{-kf}
\exp\bigl[\sum t_j \omega^{2kj} +
&
\sum \bar t_j \omega^{(2L+2-2j)k } +
2 S^2 \omega^{2k} \bigr].\cr}
}
Consider now a term $\sim \prod t_j^{\ell_j }
\bar t_j^{\bar \ell_j }(S^2)^r$ where $\ell_j = \bar \ell_j = 0$
for all but finitely many $j$.
We now apply \simpsum\ with the exponent
\eqn\simfumexp{
-f + 2 \sum j \ell_j + 2 \sum(L+1-j ) \bar \ell_j + 2 r.
}
We now must divide the problem into cases.

First suppose $N$ is odd
so  $N=2L+1$.
 In this case  $N (C^{-1})^{IJ}$ is an even integral
quadratic form, and hence the  squared-fluxes $f$ are always
even. Since $f, \ell_j,$ etc. are held fixed while
$N \rightarrow \infty$ we must have:
\eqn\oddenncond{
\eqalign{
\sum \bar \ell_j & = 0 ~ \mod ~2,\cr
-f + 2 \sum j \ell_j - 2 \sum (j-1/2)   \bar \ell_j + 2 r & = 0. \cr}
}
Thus, the large $N$ theory is summarized by:
\eqn\simpletorii{
\eqalign{
\Biggl\langle \exp\biggl( \sum t_j A_j + \sum \bar t_j \bar A_j+ I(S) \biggr)
\Biggr\rangle_f
& = \qquad\qquad  \qquad\qquad\qquad \cr
{N \over  2}  \oint {dz \over  z} z^{-f/2} e^{2 S^2 z}
\exp\bigl[\sum t_j  z^j  \bigr] \cdot
 \Biggl[ \exp \bigl[ \sum \bar t_j  z^{-j+1/2} \bigr]
&
+ \exp \bigl[ - \sum \bar t_j  z^{-j+1/2} \bigr] \Biggr]. \cr}
}
An interesting point is that {\it the above
 $1/N$ ``expansion" is exact.}

Now we consider $N$ even. The integral form
$N (C^{-1})^{IJ}$ is odd, and hence there  are two subcases
depending on whether the flux-squared $f$ is even or odd.
If $f$ is even then the evaluation proceeds as before and
\eqn\simpletoriii{
\eqalign{
\Biggl\langle \exp\biggl( \sum t_j A_j + \sum \bar t_j \bar A_j+ I(S) \biggr)
\Biggr\rangle_f
& = \qquad\qquad  \qquad\qquad\qquad \cr
{N \over  2}  \oint {dz \over  z} z^{-f/2} e^{2 S^2 z}
\exp\bigl[\sum t_j  z^j  \bigr] \cdot
 \Biggl[ \exp \bigl[ \sum \bar t_j  z^{-j } \bigr]
&
+ \exp \bigl[ - \sum \bar t_j  z^{-j } \bigr] \Biggr]. \cr}
}
If $f$ is odd then we evaluate the sums in
\simpletori\ using \oddexp.
Now the $1/N$ expansions become nontrivial and do  not
terminate.

\subsec{$X=K3$}

The situation for $X=K3$ is very similar.
$K3$ is hyperkahler so the
only basic class is $\lambda=0$. Now $\delta=2$ so
$\omega^{(N^2-1)k \delta} = \omega^{-2k}$. Thus
the results \simpletorii\simpletoriii\ continue to hold
with the simple modification:
\eqn\simpmodif{
\oint {d z \over  z} (\cdots ) \rightarrow
{\tilde \alpha} (N)^{24} {\tilde \beta}(N)^{-16} \oint {d z \over  z^2}(\cdots
).
}

\subsec{Other 4-manifolds}

For other four-manifolds with nonzero basic classes, the
evaluation of the correlators is more complicated due to the off-diagonal
couplings in \simple, which mix the different $U(1)$ factors. We will give here
some indications on the structure of these correlators in the simple case of
minimal surfaces of general type, and
sketch a possible strategy to perform a systematic large $N$ expansion.
In the case of minimal surfaces of general type, the only basic classes are
$\pm K$, where $K$ is the canonical
line bundle of the manifold,
and moreover \monopole\fandm\morganbk:
\eqn\gti{
\eqalign{
SW(-K) & = 1, \cr
SW(+K) & = (-1)^\delta. \cr}
}
We can then introduce the variables $s^I$, $I=1, \dots, N-1$, taking the values
$\pm 1$, and
define $2 \lambda^I =s^I K$. The sum over the basic classes for the $k$ vacua
now takes the
form
\eqn\sumspins{
\sum_{s^I=\pm1} (-s^I)^{\delta}  \exp \Biggl\{ -{ \pi i \over 2} K^2 \sum_{ I<J
} \tau_{IJ} s^I s^J
+  \omega^k (S,K) \sum_{I=1}^{N-1} \sin {\pi I \over  N} s^I \Biggr\}.}
This is a correlation function for a one-dimensional spin chain with $N-1$
sites, with long range
interactions given by the off-diagonal couplings $\tau_{IJ}$, and in the
presence of a
space dependent ``magnetic field" proportional to $\sin(\pi I/N)$. The large
$N$ limit of this
expression corresponds then to the thermodynamic limit of the system. This
suggests that
one can study the large $N$ limit using standard techniques in statistical
mechanics. One possible strategy is to rewrite \sumspins\ by introducing
auxiliary variables $x^I$, $I=1, \dots, N-1$, and reexpress the quadratic term
in the $s^I$ spin variables using a gaussian integral.  To do this, it is
useful to consider the invertible matrix of couplings $\tilde \tau_{IJ}$,
$I,J=1, \dots, N-1$, where the diagonal couplings come from the regular part at
the $N=1$ point
of the couplings defined in \scaling\ ({\it i.e.} after substracting the
logarithmic divergence), and the off-diagonal terms are the ones in \sumspins.
This is actually the matrix of couplings that naturally appears in the
Seiberg-Witten contribution in
\swnone. The term involving the diagonal part of $\tilde \tau_{IJ}$ is just an
overall factor depending on $N$ and $K^2$. After introducing the
auxiliary variables $x^I$, the sum over the spin variables can be easily worked
out, and the correlation function \sumspins\ becomes, for $\delta$ even:
\eqn\statint{
\eqalign{
2^{N-1}  C(N) & \int_{-\infty}^{+\infty} \prod_{I=1}^{N-1} dx^I  \exp \Biggl\{
-{i \over \pi K^2} \sum_{I,J} \tilde \tau_{IJ}^{-1}  x^I x^J \cr
&\,\,\,\ + \sum_{I=1}^{N-1}  \log \cosh
  \biggl[ \omega^k (S,K) \sin {\pi I \over  N} +  x^I \biggr] \Biggr\},\cr}}
where $C(N)$ is an overall constant depending on $N$ and $K^2$. An analogous
expression involving $\sinh$ can be obtained for $\delta$ odd. Notice that, for
minimal surfaces of general
type one has $K^2>0$. One could evaluate this integral using a saddle-point
approximation (which should be a good one in the large $N$ limit) and
then systematically computing the corrections to the saddle-point.

For other four-manifolds, the set of basic classes is more complicated, but in
the case
of algebraic surfaces one has a complete description of this set and the
Seiberg-Witten invariants have been explicitly computed \monopole\fandm. The
structure of \simple\ indicates that the contribution of
an ${\cal N}=1$ vacuum will be given again by a correlation function in some
statistical mechanical system.  As we have suggested, this analogy could prove
useful in studying the properties of the Donaldson-Witten
function in the large $N$ limit, and one may likely find interesting phenomena
(like phase transitions.)
Another reason to study the large $N$ limit of Donaldson-Witten theory is a
possible relation to
topological strings, as   was conjectured in \icm\ (a relation between the
large $N$ limit of
Chern-Simons topological gauge theory and topological strings has been recently
found in \gv.)

\subsec{Possible applications to Dbranes and matrix theory}

There are several ways in which $\CN=2$ SYM
is realized in the context of string theory, D-branes
and M-theory. See \klemmrev\  and \giveonkut\ for
recent reviews. We limit ourselves here to a few
brief and superficial remarks.

Perhaps the most direct applications are to
matrix theory. As noted above the physical theory
and twisted theory correlators coincide
for the hyperkahler 4-manifolds $X=T^4,K3$.
The correlators on the 4-torus can be
interpreted as  a kind of
finite temperature partition function:
$Z_{DW}= {\Tr}e^{-\beta H} (-1)^F  \exp[\sum t_i  A_i]  $.

In \susskind\ a matrix theory approach to studying
Schwarzschild black holes was proposed. This
approach requires the existence of a singularity in
the equation of state, which should translate into
a singularity in ${\Tr } e^{-\beta H}  \prod \CO_i $
at large $N$ as a function of $\beta$.
Although we are studying theories with 8 rather
than 16 supercharges, one would expect the
phenomenon required by \susskind\ to be rather
generic. Unfortunately
we find no evidence of the discontinuity in
$N$ posited in  \susskind. This might be due to the
insertion of $(-1)^F$, and indeed that
is consistent with the discussion in
section 3.2 of \holography.

In the realization
of Seiberg-Witten  theory via  $M$-theory 5-branes
described in
\wittmfive\giveonkut\ the correlators
of ${\Tr} \phi^{2n}$ carry information about the quantum
distribution of positions of $(D4)$ branes. The
tendency of large $N$ correlators to vanish as
described above   would seem to
suggest that if the 5branes are wrapped in the
$x^{1,2,3}$ directions (to use the standard choice
of coordinates in  \wittmfive\giveonkut ) at finite
temperature then the D4 branes - or tubes between
NS5 branes - are very uniformly distributed.

When the 4-manifold
 $X$ is not hyperkahler then the twisted and
physical correlators can differ. However, topological
correlators can very well be relevant in the theory
of D-branes \BSV\
so the   above results might
also find   applications in the theory of branes in
more complicated compactifications of string/M theory.

\bigskip
\centerline{\bf Acknowledgements}\nobreak
\bigskip

We would like to thank E. Witten for several
important discussions and remarks. We also thank
M. Douglas, J. Harvey, and D. Freedman for some
useful remarks.
Many of these results were announced in
\mm\ and in a talk   at the  CERN workshop
on Non-perturbative Aspects of Strings, Branes and Fields,
Dec. 8-12. GM thanks the organizers
Luis \'Alvarez-Gaum\'e, Costas Kounnas, and Boris Pioline
for the opportunity to present these results.
This work  is supported by
DOE grant DE-FG02-92ER40704.

\listrefs

\bye